\documentclass[twocolumn,aps]{revtex4}

\pdfoutput=1

\usepackage[english]{babel}

\usepackage{dcolumn}

\usepackage{amsmath}
\usepackage{graphicx}
\usepackage{subfigure}
\usepackage{multirow}
\usepackage{txfonts}
\def\bea{\begin{eqnarray}}
\def\eea{\end{eqnarray}}
\def\be{\begin{equation}}
\def\ee{\end{equation}}

\begin{document}

\title{Gaussian processes reconstruction of dark energy from observational data}

\author{Ming-Jian Zhang$^{a}$}
\author{Hong Li$^{a}$}\email[Corresponding author: ]{hongli@ihep.ac.cn}
\affiliation{$^a$Key Laboratory of Particle Astrophysics, Institute of High Energy Physics,
Chinese Academy of Science, P. O. Box 918-3, Beijing 100049, China}

\begin{abstract}
In the present paper, we investigate the dark energy equation of state using the Gaussian processes analysis method, without confining a particular parametrization. The reconstruction is carried out by adopting the background data including supernova and Hubble parameter, and perturbation data from the growth rate. It suggests that the background and perturbation data both present a hint of dynamical dark energy. However, the perturbation data have a more promising potential to distinguish non-evolution dark energy including the cosmological constant model. We also test the influence of some parameters on the reconstruction. We find that the matter density parameter $\Omega_{m0}$ has a slight effect on the background data reconstruction, but has a notable influence on the perturbation data reconstruction. While the Hubble constant presents a significant influence on the reconstruction from background data.

\end{abstract}

\maketitle

\section{Introduction}
\label{introduction}

Multiple experiments have consistently approved the cosmic late-time accelerating expansion. Observations contributing to this pioneering discovery contain the type Ia supernova (SNIa) \citep{riess1998supernova,perlmutter1999measurements}, large scale structure \citep{tegmark2004cosmological}, cosmic microwave background (CMB)
anisotropies  \citep{spergel2003wmap}, and baryon
acoustic oscillation (BAO) peaks \citep{eisenstein2005detection}. Theoretical paradigms trying to explain this discovery include the exotic dark energy with repulsive gravity, or modification to general relativity \citep{barrow1988inflation,dvali20004d}, or violation of cosmological principle \citep{lemaitre1933univers,tolman1934effect,bondi1947spherically}. In which, dark energy theory attracts lots of interests. For understanding the nature of dark energy, a crucial parameter is the equation of state (EoS) $w$, which is the ratio of pressure to energy density. Basing on the value of $w$, dark energy can be classified to different categories. The cosmological constant model with  $w=-1$ is the most notable candidate. In addition to this one, the time evolution model, Chevallier-Polarski-Linder (CPL) \cite{chevallier2001accelerating,linder2003exploring}  is also a potential competitor. A short review can be seen in Ref.  \cite{yang2017latest}.

However, above understanding of the dark energy is a parametrization on  $w$. It is, after all, an ansatz of the dark energy. To extract the information of EoS honestly, Huterer and Starkman \cite{huterer2002parameterization} first proposed the principal component analysis technique in dark energy study. It is a model-independent way which treats the $w$ as a piecewise constant in each redshift bin. By extracting essential information from multiple observational data, one can obtain a series of orthogonal eigenfunctions to expand the EoS $w$. In following Refs. \cite{clarkson2010direct,huterer2005uncorrelated,nesseris2014comparison,zheng2017constraints}, this method was greatly adopted and improved in different forms.

Another effective technique, Gaussian processes (GP), is also model-independent. Unlike the parametrization constraint, this approach does not rely on any artificial dark energy template. It can reconstruct the $w$ directly via its relationship with the observational variable. In this process, it firstly assumes that each observational data satisfies a Gaussian distribution. Thus, the observational data should satisfy a multivariate normal distribution. Relationship between two different data points is connected by a covariance function. Using this covariance function, values of data at other redshift points which have not be observed also can be obtained because they all obey this probability distribution. Moreover, derivative of these data also can be calculated using the covariance function. Finally, with the preparation of more data, a variable or goal function can be reconstructed at any redshift point via their relationship with the data and its derivatives. We note that the primary task in this Gaussian processes is to determine the covariance function at different redshift points using the observational data. Moreover, determination of the covariance function has nothing to do with the $w$. Thus, its understanding on the $w$ is more faithful. In cosmology, it has incurred a wide application in reconstructing dark energy \cite{holsclaw2010nonparametric,seikel2012reconstruction} and cosmography \cite{shafieloo2012gaussian}, or testing standard concordance model \cite{yahya2014null} and distance duality relation \cite{santos2015two}, or determinating the interaction between dark matter and energy \cite{yang2015reconstructing} and spatial curvature \cite{cai2016null}. Also, this method has been used to study the dark energy \cite{seikel2012reconstruction,seikel2012using,wang2017improved}. However, most focus were on the background observational data, such as supernova and Hubble parameter data. Moreover, they did not consider the effect of matter density parameter.

In the present paper, we want to learn more about the dark energy. The data we use are not only the background data including supernova luminosity distance and Hubble parameter, but also the perturbation data, growth rate of structure $f\sigma_8$. Moreover, we consider to test the influence of some parameters including the matter density. For the perturbation level data, they measure the redshift-space distortions. It has been evidenced that these data can provide tight constraint on the parameter space \cite{2012PhLB..717..299S,2014PhRvD..89d3511Y}, or test the cosmic acceleration \cite{2008Natur.451..541G}, or distinguish the Galileon model from $\Lambda$CDM model \cite{2011PhRvD..83d3515D,2012JCAP...08..026A}, or distinguish some modified gravity models \cite{2017arXiv170508797B}. Motivated by the advantage of growth rate data, we expect to obtain a new model-independent constraint on the dark energy. Another difference from previous work is that the SNIa and $H(z)$ data here are used as a combination of background data, not two single ones.

This paper is organized as follows: In Section~\ref{methodology}, we introduce some theoretical basis and the GP approach. And in Section~\ref{data} we introduce the relevant data we use. We present the reconstruction result in Section \ref{result}. Finally, in Section \ref{conclusion} conclusion and discussion are drawn.

\section{Methodology}
\label{methodology}

In this section, we introduce some theoretical basis and the GP approach.

\subsection{Theoretical basis}  \label{basis}

\textit{On background level}: In the Friedmann-Robertson-Walker universe, the luminosity distance function $d_L(z)$ of SNIa is
\begin{equation}
    \label{dL:define}
    d_L(z) =  \frac{c}{H_0} (1+z)  \int^z_0 \frac{ H_0 \mathrm{d}
    z'}{H(z')} .
\end{equation}
In the GP reconstruction, it is very convenient to define a dimensionless comoving luminosity distance
\be   \label{D:define}
D(z) \equiv \frac{H_0}{c} \frac{d_L (z)}{1+z} .
\ee
Obviously, combining Eq. \eqref{D:define} and \eqref{dL:define}, taking derivative with respect to redshift $z$, it is easy for us to obtain the relation between Hubble parameter and distance $D(z)$
\be  \label{Hubble_SN}
E (z) \equiv \frac{H(z)}{H_0} = \frac{ 1}{D'} ,
\ee
where $E(z)$ is the dimensionless Hubble parameter, and the prime denotes derivative with respect to redshift $z$.

\textit{On perturbation level}: In the general relativity and a background universe filled with matter and unclustered dark energy, the evolution of matter density contrast, $\delta(z) \equiv \frac{\delta \rho_m}{\rho_m} (z)$, at scales much smaller than the Hubble radius should obey the following second order differential equation
\begin{equation}  \label{eq:delta}
   \ddot{\delta} + 2H \dot{\delta} -4\pi G \rho_m \delta =0 ,
\end{equation}
where $\rho_m$ is the background matter density, $\delta \rho_m$ represents its first-order
perturbation, the dot denotes derivative with respect to cosmic time $t$.
Basing the relation $d/dt = a H (d/d a)$, we can change the argument of Eq. \eqref{eq:delta} from cosmic time to scale factor. Subsequently, according to the relation between scale factor and redshift, Hubble parameter in Eq. \eqref{eq:delta} can be expressed as an integral over the perturbation and its derivative  \cite{starobinsky1998determine,chiba2007consistency}
\be  \label{solution:E_delta}
E^2(z) = 3 \Omega_{m0} \frac{(1+z)^2}{\delta '(z) ^2 } \int_z^{\infty} \frac{\delta}{1+z} (-\delta ') d z ,
\ee
where $\Omega_{m0}$ is the matter density parameter today and the prime denotes derivative with respect to redshift $z$. Conversely, the solution of perturbation also can be solved as an integral of Hubble parameter. From the Eq. \eqref{solution:E_delta}, we find that the Hubble parameter $E^2(z)$ tends to zero when the redshift in integral $z \rightarrow \infty$. When the redshift $z=0$, we have the initial condition
\be  \label{initial_condition}
1 = \frac{3 \Omega_{m0}}{\delta '(z=0) ^2} \int_0^{\infty} \frac{\delta}{1+z} (-\delta ') d z .
\ee
For the integral in Eq. \eqref{solution:E_delta}, it usually can be calculated by the relation $\int_z^{\infty} f(z) dz = \int_0^{\infty} f(z) dz - \int_0^{z} f(z) dz$. In previous work, they usually substitute the matter density parameter $\Omega_{m0}$ in Eq. \eqref{initial_condition} into Eq. \eqref{solution:E_delta}. The Hubble parameter is, thus, expressed as a ratio of two integrals. However, perturbation $\delta$ at higher redshift $z \gtrsim 5$ maybe cannot be determined from observation \cite{starobinsky1998determine,chiba2007consistency}. Therefore, in practice, it may be difficult to calculate the integral $\int_0^{\infty} \frac{\delta}{1+z} (-\delta ') d z$. In the present paper, we deal with it in a diametrically opposite way. That is, we replace the integral $\int_0^{\infty} \frac{\delta}{1+z} (-\delta ') d z$ using the parameter $\Omega_{m0}$. Consequently, Eq. \eqref{solution:E_delta} can be written as
\be  \label{Hubble_delta}
E^2(z) = (1+z)^2  \frac{\delta '(z=0) ^2}{\delta '(z) ^2 }  -  3\Omega_{m0} \frac{(1+z)^2}{\delta '(z) ^2 } \int_0^{z} \frac{\delta}{1+z} (-\delta ') d z .
\ee

Observationally, current cosmological surveys cannot provide direct measurement of perturbation $\delta(z)$, but can provide a related observation, the growth rate measurement $f\sigma_8$ from redshift-space distortions (RSD) caused by the peculiar motions of galaxies \cite{1987MNRAS.227....1K}. Here, the growth rate $f$ is defined by the derivative of the logarithm of perturbation $\delta$ with respect to logarithm of the cosmic scale
\be  \label{f:define}
f \equiv \frac{d \, \textmd{ln} \delta}{d \, \textmd{ln} a}  = -(1+z) \frac{d \, \textmd{ln} \delta}{d \, z} = -(1+z) \frac{\delta '}{\delta} .
\ee
While the function
\be
\sigma_8 (z) = \sigma_8 (z=0) \frac{\delta (z)}{\delta (z=0)}
\ee
is the linear theory root-mean-square mass fluctuation within a sphere of radius $8h^{-1}$Mpc, where $h$ is the dimensionless Hubble constant. Because RSD measurements are sensitive to the product of these two functions, they have been used in a wide range to constrain the evolution of universe and directly test the general relativity, thereby providing an insight on the fundamental physics.

In the light of above two definitions, the growth rate of structure is written as
\be
f \sigma_8 = -\frac{\sigma_8 (z=0)}{\delta (z=0)} (1+z) \delta ' .
\ee
It is easy for us to have
\be  \label{Eq:dp}
\delta ' = - \frac{\delta (z=0)}{\sigma_8 (z=0)} \frac{f \sigma_8}{1+z}  .
\ee
Obviously, derivative of the perturbation $\delta$ can be easily transferred or reconstructed from the observational RSD data. Taking an integral to the two sides of Eq. \eqref{Eq:dp} over redshift, we have
\be  \label{delta_end}
\delta  = \delta (z=0) - \frac{\delta (z=0)}{\sigma_8 (z=0)} \int_0^{z} \frac{f \sigma_8}{1+z} dz .
\ee
For the constant $\delta (z=0)$, it was commonly fixed as the normalization value $\delta (z=0)=1$ \cite{2017JCAP...08..008G,2016arXiv161202484C} or a fiducial value. In this paper, we would like to test the influence of different $\delta(z=0)$ on the reconstruction, at the normalization value $\delta(z=0)=1$ and a fiducial value $\delta(z=0)=0.7837$. We also intend to test the influence of different $\sigma_8 (z=0)$ on the reconstruction.

We consider a spatial flat Friedmann-Robertson-Walker universe with dark matter and dark energy
\be  \label{DE_model}
E^2(z) = \Omega_{m0} (1+z)^3 + (1-\Omega_{m0}) \exp \left[3 \int^z_0 \frac{1+w(z')}{1+z'} dz' \right] .
\ee
Obviously, the EoS of dark energy can be obtained by taking the derivative with respect to $z$ on the two sides of above equation. Substituting Eq.  \eqref{Hubble_SN} and \eqref{Hubble_delta} into \eqref{DE_model} respectively, we have
\be   \label{EoS:SN}
w (z) = \frac{1}{3} \frac{-2 (1+z) D'' - 3D'}{D' - \Omega_{m0} (1+z)^3 D^{'3}}
\ee
for the background data, and
\be   \label{EoS:delta}
w (z) = \frac{1}{3} \frac{(1+z) E^2 (z) ' -3 E^2(z) }{E^2 (z) - \Omega_{m0} (1+z)^3}
\ee
for the RSD data.

\begin{figure}
\centering
\includegraphics[width=7cm,height=5.5cm]{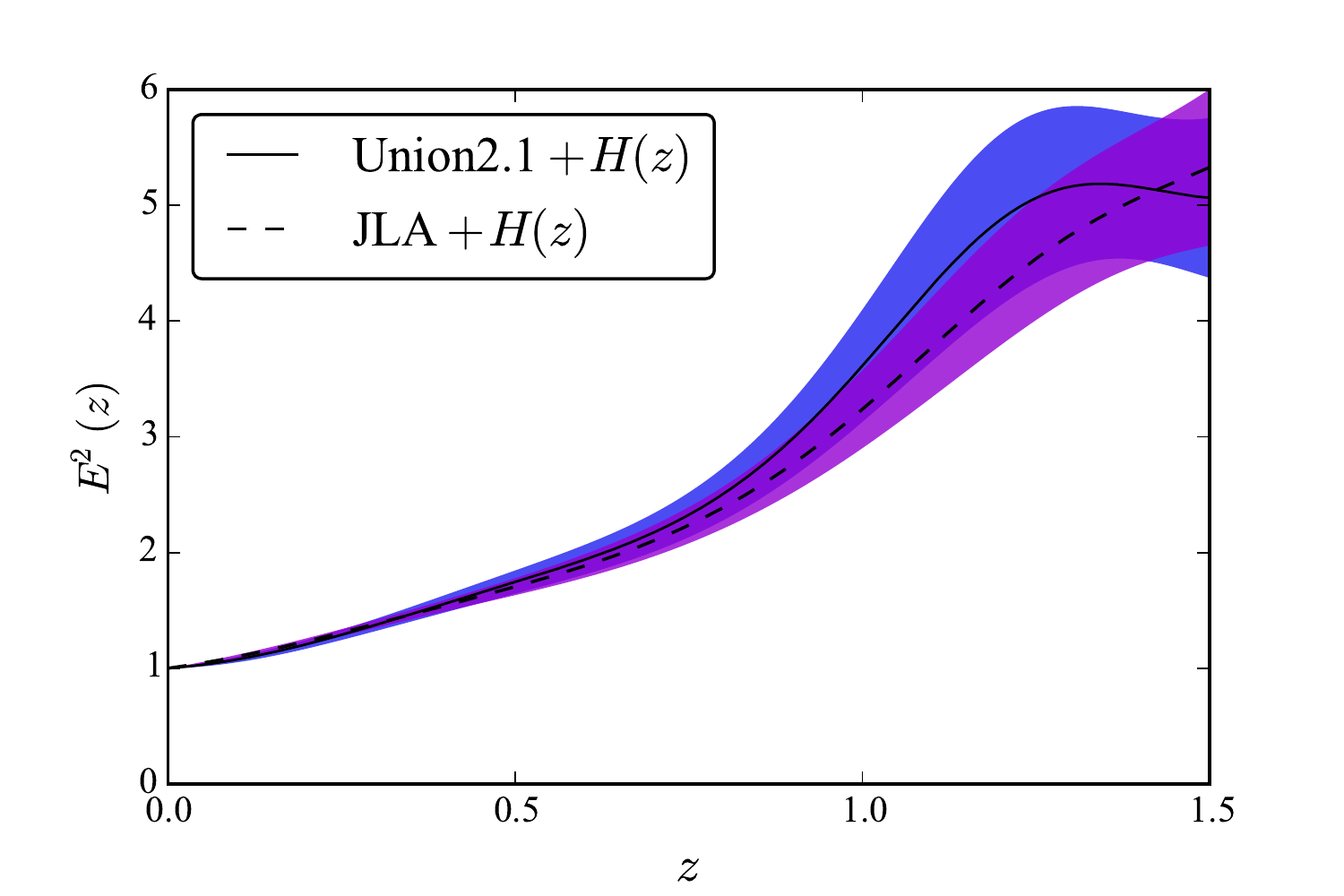}
\includegraphics[width=7cm,height=5.5cm]{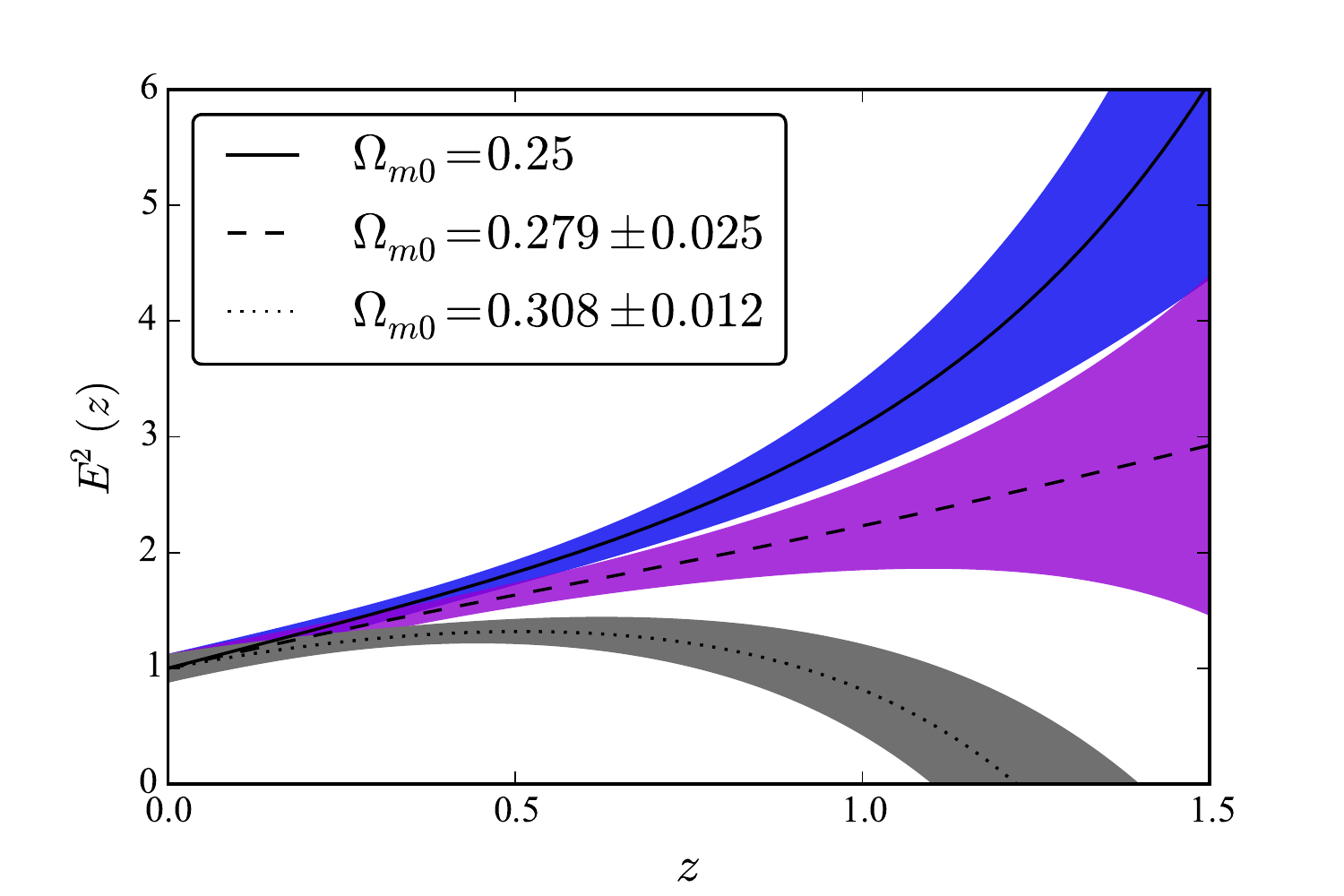}
    \caption{\label{Fig:E2} The square of Hubble parameter from the background data (\textit{upper panel}) and perturbation data (\textit{lower panel}).  }
\end{figure}

\subsection{Gaussian processes}  \label{GP}
\begin{figure*}
\centering
\includegraphics[width=18cm,height=5cm]{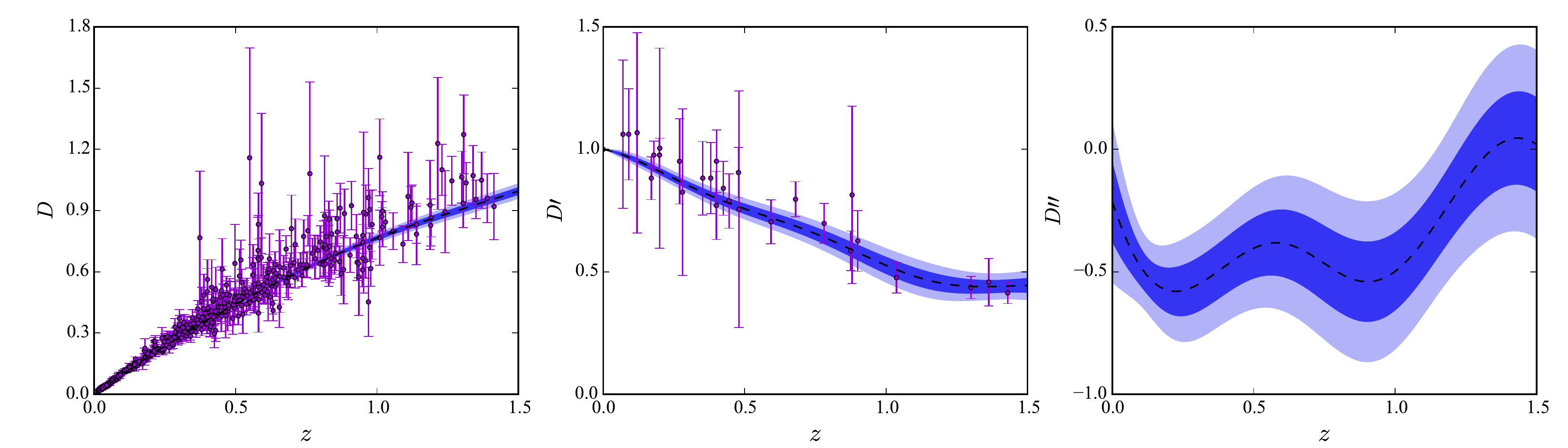}
    \caption{\label{union21_D} The reconstruction of distance $D$ and its derivatives using the combination of Union2.1 and $H(z)$ data. Data with errorbars in the left panel are observational supernova data. Data with errorbars in the middle panel are $H(z)$ data. }
\end{figure*}

To reconstruct the goal function $f(z)$, a parametrization constraint or a model-independent technique should be carried out. For the former method, a prior form on the constrained function $f(z)$ is usually restricted. For example, to understand the dark energy, EoS $w(z)$ is assumed to be the CPL model with two artificial parameters $w_0$ and $w_a$. Instead, a model-independent method such as the Gaussian processes, does not limit to a particular parametrization form. It only needs a probability on the goal function $f(z)$. A Gaussian process is a generalization of the Gaussian probability distribution. Assuming the observational data, such as the distance $D$, obeys a Gaussian distribution with mean and variance, the posterior distribution of goal function $f(z)$ can be expressed via the joint Gaussian distribution of different data of distance $D$. In this process, the key ingredient is the covariance function $k(z, \tilde{z})$ which correlates the values of different distance $D(z)$ at points $z$ and $\tilde{z}$. Commonly, the covariance function $k(z, \tilde{z})$ has several types, and most associated with two hyperparameters $\sigma_f$ and $\ell$ which can be determined by the observational data via a marginal likelihood. With the trained covariance function, the data can be extended to any redshift points. Using the relation between the goal function $f(x)$ and distance data $D$, the goal function can be reconstructed. Due to its model-independence, this method has been widely applied in the reconstruction of dark energy EoS \cite{holsclaw2010nonparametric1,holsclaw2010nonparametric,seikel2012reconstruction}, or in the test of the concordance model \cite{seikel2012using,yahya2014null}, or determination to the dynamics of dark energy by dodging the matter degeneracy \cite{busti2016dodging}.

For the covariance function $k(z, \tilde{z})$, many templates are available. The usual choice is the squared exponential $k(z, \tilde{z}) = \sigma_f^2 \exp[-|z-\tilde{z}|^2 / (2 \ell^2)]$.  Analysis in Ref. \cite{seikel2013optimising} shows that the Mat\'ern ($\nu=9/2$) covariance function is a better choice to present suitable and stable result. It thus has been widely used in previous work \cite{yahya2014null,yang2015reconstructing}. It is read as
\begin{eqnarray}
k(z,\tilde{z}) &=& \sigma_f^2
  \exp\Big(-\frac{3\,|z-\tilde{z}|}{\ell}\Big) \nonumber \\
  &&~\times \Big[1 +
  \frac{3\,|z-\tilde{z}|}{\ell} + \frac{27(z-\tilde{z})^2}{7\ell^2}  \nonumber\\
&&~~~~~~
+ \frac{18\,|z-\tilde{z}|^3}{7\ell^3} +
  \frac{27(z-\tilde{z})^4}{35\ell^4} \Big]. \label{mat}
\end{eqnarray}
With the chosen Mat\'ern ($\nu=9/2$) covariance function, we can reconstruct the EoS of dark energy by modifying the publicly available package GaPP developed by \citet{seikel2012reconstruction}. We refer the reader to Ref. \cite{seikel2012reconstruction} for more details on the GP method.

\section{Observational data}
\label{data}
\begin{figure*}
\centering
    \includegraphics[width=5.9cm,height=5.4cm]{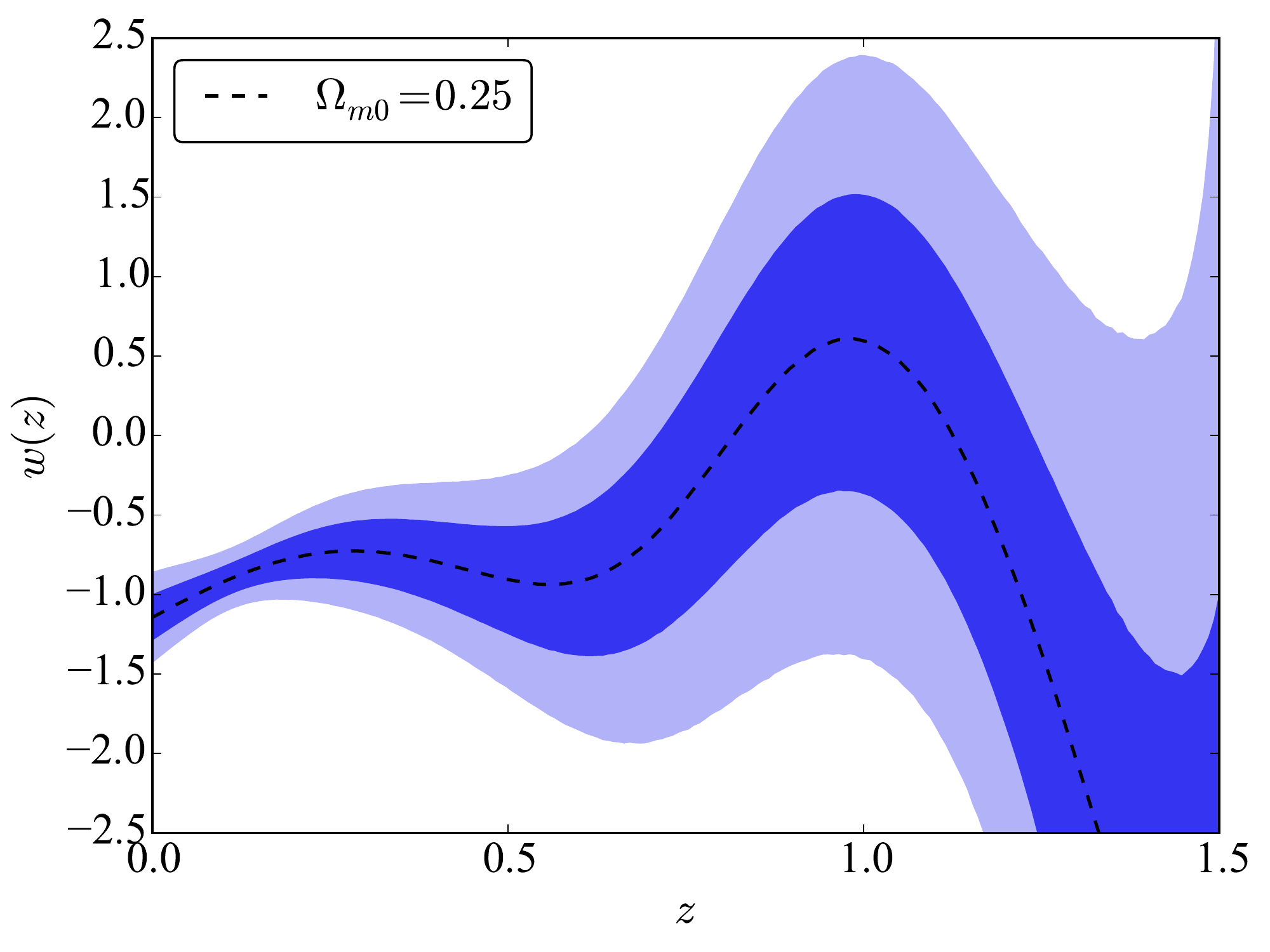}
    \includegraphics[width=5.9cm,height=5.4cm]{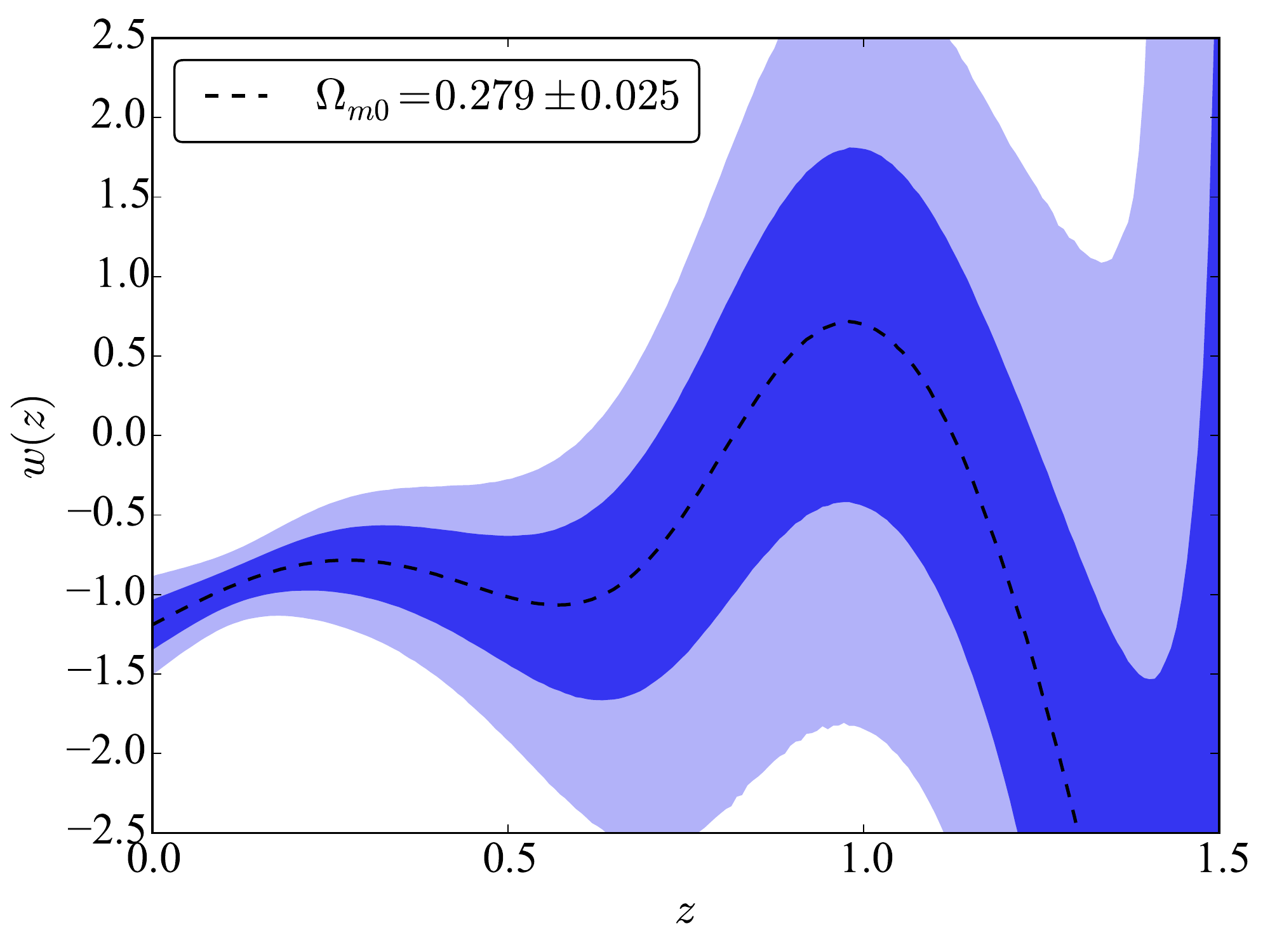}
    \includegraphics[width=5.9cm,height=5.4cm]{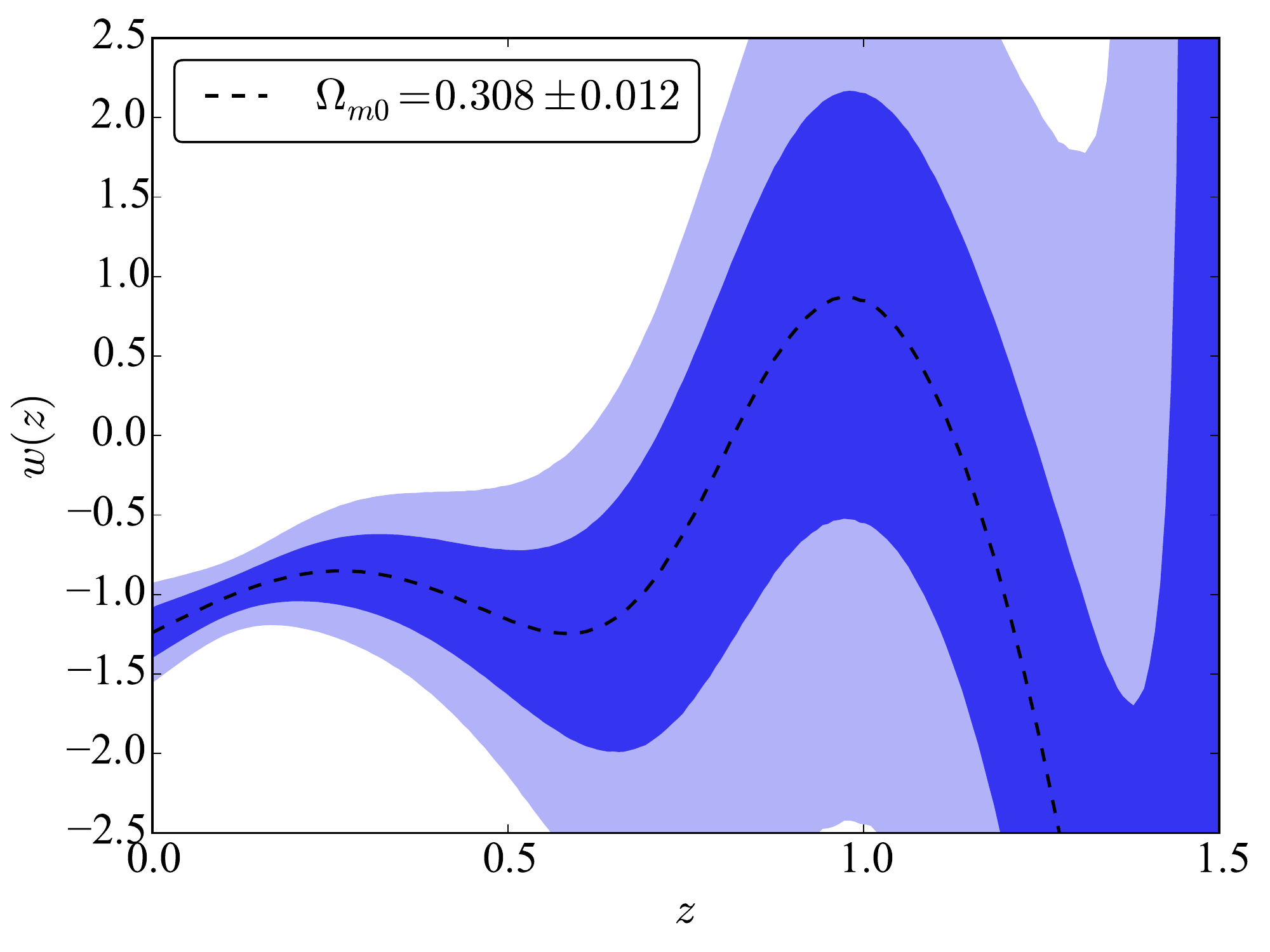}
    \caption{\label{union21_w} Dark energy reconstruction for different matter density parameter using the combination of Union2.1 and $H(z)$ data.  }
\end{figure*}

In this section, we report the related observational data.

For the supernova data, we use the Union2.1 compilations \cite{suzuki2012hubble} released by the Hubble Space Telescope Supernova Cosmology Project and the JLA datasets \cite{2014A&A...568A..22B}. Usually, they are presented as tabulated distance modulus with errors. For the Union2.1 data, they contain 580 dataset. Their redshift regions are able to span over $z<1.414$. For the JLA sample, it spans a range at redshift $0.01 <z <1.3$. It consists of 740 SNIa datasets, including three-season data from SDSS-II ($0.05 < z <0.4$), three-year data from SNLS ($0.2 < z <1$), HST data ($0.8 < z <1.4$), and several low-redshift samples ($z <0.1$). According to their test, the binned JLA data have a same constraint power as the full version of the JLA likelihood on the cosmological model. In our calculation, we use the 31 binned distance modulus with covariance matrix, which is issued in their Ref. \cite{2014A&A...568A..22B}.

The distance modulus of each supernova can be estimated as
    \begin{equation}  \label{mu:define}
    \mu (z) =  5 \textrm{log}_{10}d_L(z)+25,
    \end{equation}
where $d_L$ is the luminosity distance in Eq. \eqref{dL:define}.
In our calculation, we set the same prior of $H_0$ as the following $H(z)$ data and include the covariance matrix with systematic errors in our calculation. To obtain the dimensionless comoving luminosity distance $D(z)$, we should make a transformation from the distance modulus via Eq. \eqref{D:define}. Moreover, the theoretical initial conditions $D(z=0)=0$ and $D'(z=0)=1$ are also taken into account in the calculation.

\begin{table}[]
\caption{A compilation of recent RSD data from different surveys.
\label{tab:fsig8_data}}
\centering
\renewcommand\arraystretch{1.15}
\begin{tabular}{cccr@{$\pm$}lcc}

\hline
Index & Dataset & $z$ & \multicolumn{2}{c}{$f\sigma_8(z)$}  &Refs. & Year  \\
\hline
\hline
1 & THF         & 0.020  & 0.360&0.0405   & \cite{hudson2012growth} & 2013  \\

2 & 6dFGRS     & 0.067  & 0.423&0.055    & \cite{beutler20126df}  & 2012  \\

3 &SDSS-veloc  & 0.100  & 0.370&0.130     & \cite{feix2015growth}  & 2015 \\

4 & 2dFGRS     & 0.170  & 0.510&0.060    & \cite{song2009reconstructing} & 2009 \\

5 &  WiggleZ   & 0.220  & 0.420&0.070    & \cite{blake2011wigglez}       & 2011 \\

6 &SDSS-LRG-200 & 0.250  & 0.3512&0.0583 & \cite{samushia2012interpreting}  & 2012 \\

7 &SDSS-BOSS   & 0.300  & 0.407&0.055    & \cite{tojeiro2012clustering}  & 2012 \\

8 &SDSS-BOSS DR12   & 0.310  & 0.384&0.083    & \cite{2017arXiv170905173W}  & 2017 \\

9 &BOSS-LOWZ   & 0.320  & 0.384&0.095    & \cite{sanchez2014clustering}  & 2013 \\

10 &SDSS-LRG    & 0.350  & 0.440&0.050    & \cite{song2009reconstructing}  & 2009 \\

11 &SDSS-BOSS DR12   & 0.360  & 0.409&0.098    & \cite{2017arXiv170905173W}  & 2017 \\

12&SDSS-LRG-200 & 0.370  & 0.4602&0.0378  & \cite{samushia2012interpreting}  & 2011 \\

13& GAMA       & 0.380  & 0.440&0.060    & \cite{blake2013galaxy}  & 2013 \\

14&SDSS-BOSS   & 0.400  & 0.419&0.041    & \cite{tojeiro2012clustering}  & 2012 \\

15& WiggleZ    & 0.410  & 0.450&0.040    & \cite{blake2011wigglez}  & 2011 \\

16&SDSS-BOSS DR12   & 0.440  & 0.426&0.062    & \cite{2017arXiv170905173W}  & 2017 \\

17&SDSS-BOSS DR12   & 0.480  & 0.458&0.063    & \cite{2017arXiv170905173W}  & 2017 \\

18&SDSS-BOSS   & 0.500  & 0.427&0.043    & \cite{tojeiro2012clustering}  & 2012 \\

19&BOSS DR12   & 0.510  & 0.458&0.038    & \cite{2017MNRAS.470.2617A}  & 2016 \\

20&SDSS-BOSS DR12   & 0.520  & 0.483&0.075    & \cite{2017arXiv170905173W}  & 2017 \\

21&SDSS-BOSS DR12   & 0.560  & 0.472&0.063    & \cite{2017arXiv170905173W}  & 2017 \\

22&SDSS-LRG-200     & 0.570  & 0.423&0.052    & \cite{samushia2014clustering}  & 2014 \\

23&SDSS-BOSS DR12   & 0.590  & 0.452&0.061    & \cite{2017arXiv170905173W}  & 2017 \\

24&SDSS-BOSS   & 0.600  & 0.433&0.067    & \cite{tojeiro2012clustering}  & 2012 \\

25&BOSS DR12   & 0.610  & 0.436&0.034    & \cite{2017MNRAS.470.2617A}  & 2016 \\

26&SDSS-BOSS DR12   & 0.640  & 0.379&0.054    & \cite{2017arXiv170905173W}  & 2017 \\

27&WiggleZ     & 0.730  & 0.437&0.072    & \cite{blake2012wigglez}  & 2012 \\

28& VVDS       & 0.770  & 0.490&0.018    & \cite{song2009reconstructing}  & 2009 \\

29& Vipers     & 0.800  & 0.470&0.080    & \cite{de2013vimos}  & 2013 \\

30&Vipers PDR-2 & 0.860  & 0.400&0.110    & \cite{pezzotta2017vimos}  & 2016 \\

31& eBOSS DR14 & 0.978  &  0.379&0.176   & \cite{2018arXiv180103043Z} & 2018 \\

32&Vipers v7   & 1.050  & 0.280&0.080    & \cite{Wilson2016thesis}  & 2016 \\

33& eBOSS DR14 & 1.230  &  0.385&0.099   & \cite{2018arXiv180103043Z}  & 2018 \\

34& eBOSS DR14 & 1.526  &  0.342&0.070   & \cite{2018arXiv180103043Z}  & 2018 \\

35& eBOSS DR14 & 1.944  &  0.364&0.106   & \cite{2018arXiv180103043Z}  & 2018 \\
\hline
\hline
\end{tabular}
\end{table}

For the $H(z)$ data, they were not direct products from a tailored telescope, but can be acquired via two ways. One is to calculate the differential ages of galaxies
\cite{jimenez2008constraining,simon2005constraints,stern2010cosmic}, usually called cosmic chronometer. The other is the deduction from the BAO peaks in the galaxy power spectrum
\cite{gaztanaga2009clustering,moresco2012improved} or from the BAO
peak using the Ly$\alpha$ forest of QSOs \cite{delubac2013baryon}. In the present paper, we use the 30 cosmic chronometer data points which compiled in Table 1 of Ref. \cite{zhang2016test}, because the latter method is model-dependent. An underlying cosmology is needed to calculate the sound horizon in the latter method. After the preparation of $H(z)$ data, we should normalize them to obtain the dimensionless one $E(z)=H(z)/H_0$. Obviously, the initial condition $E(z=0)=1$ should be taken into account in our calculation. Considering the error of Hubble constant, we can calculate the uncertainty of $E(z)$
\be
\sigma_E^2 = \frac{\sigma_H^2}{H_0^2} + \frac{H^2}{H_0^4} \sigma_{H_0}^2 .
\ee
We utilize the same prior of $H_0$ as the supernova data. Different from previous most work, we do not use the $H(z)$ data alone. We combine them with the supernova data as a derivative of distance $D$, using the relation
$
D' = \frac{1}{E(z)}.
$
To test the influence of Hubble constant,  we respectively consider two priors on the $w$ reconstruction, namely, $H_0=73.24 \pm 1.74$ km s$^{-1}$Mpc$^{-1}$ with 2.4\% uncertainty \cite{riess20162} and  $H_0=71.00 \pm 2.80$ km s$^{-1}$Mpc$^{-1}$ from the latest determination \cite{doi:10.1093/mnras/stx2710}.

To probe the growth of structure, several promising types of cosmological measurements were proposed, such as the clustering of galaxies in spectroscopic surveys, counts of galaxy clusters, and weak gravitational
lensing.

\begin{figure*}
\centering
    \includegraphics[width=5.9cm,height=5.4cm]{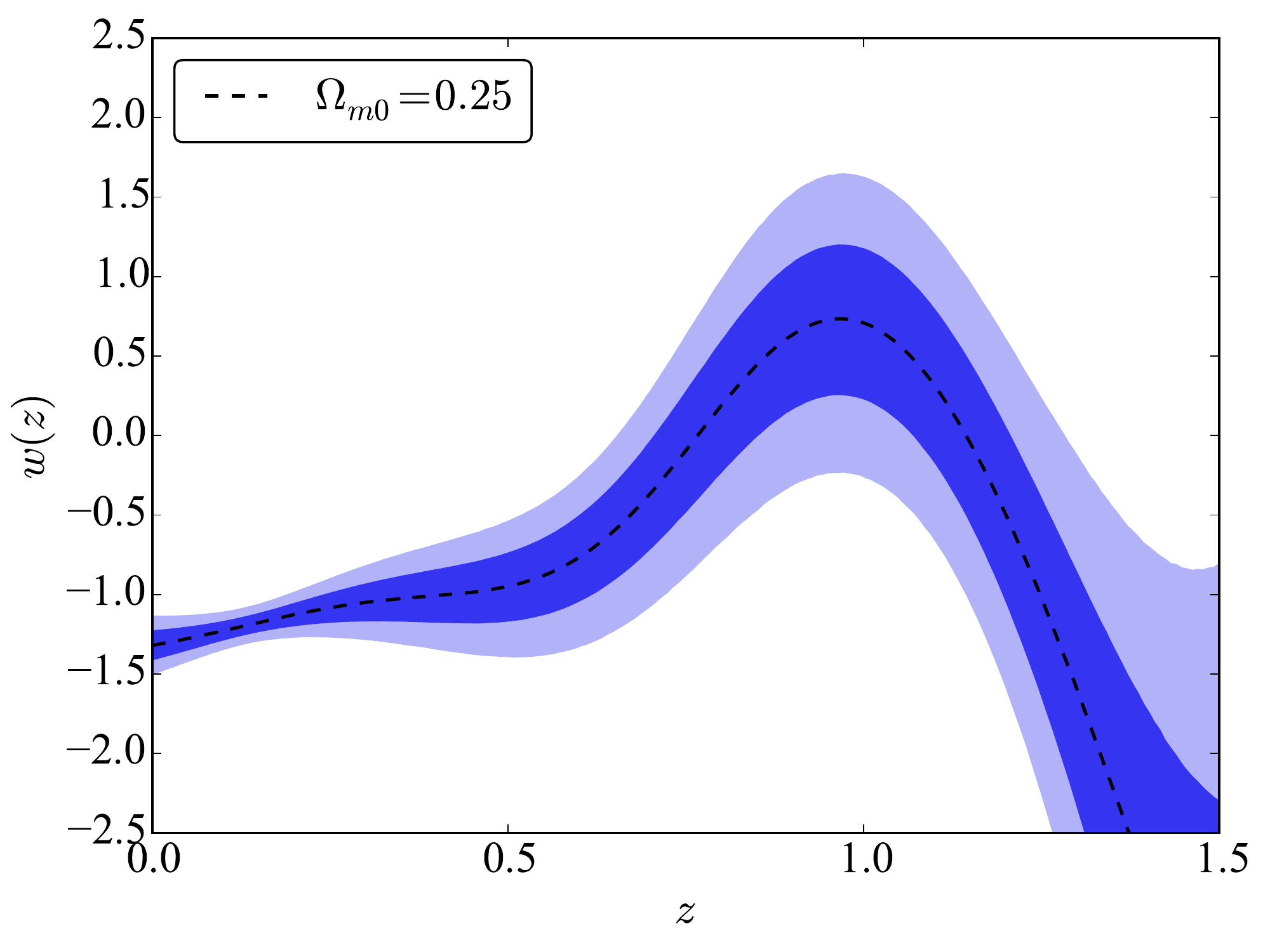}
    \includegraphics[width=5.9cm,height=5.4cm]{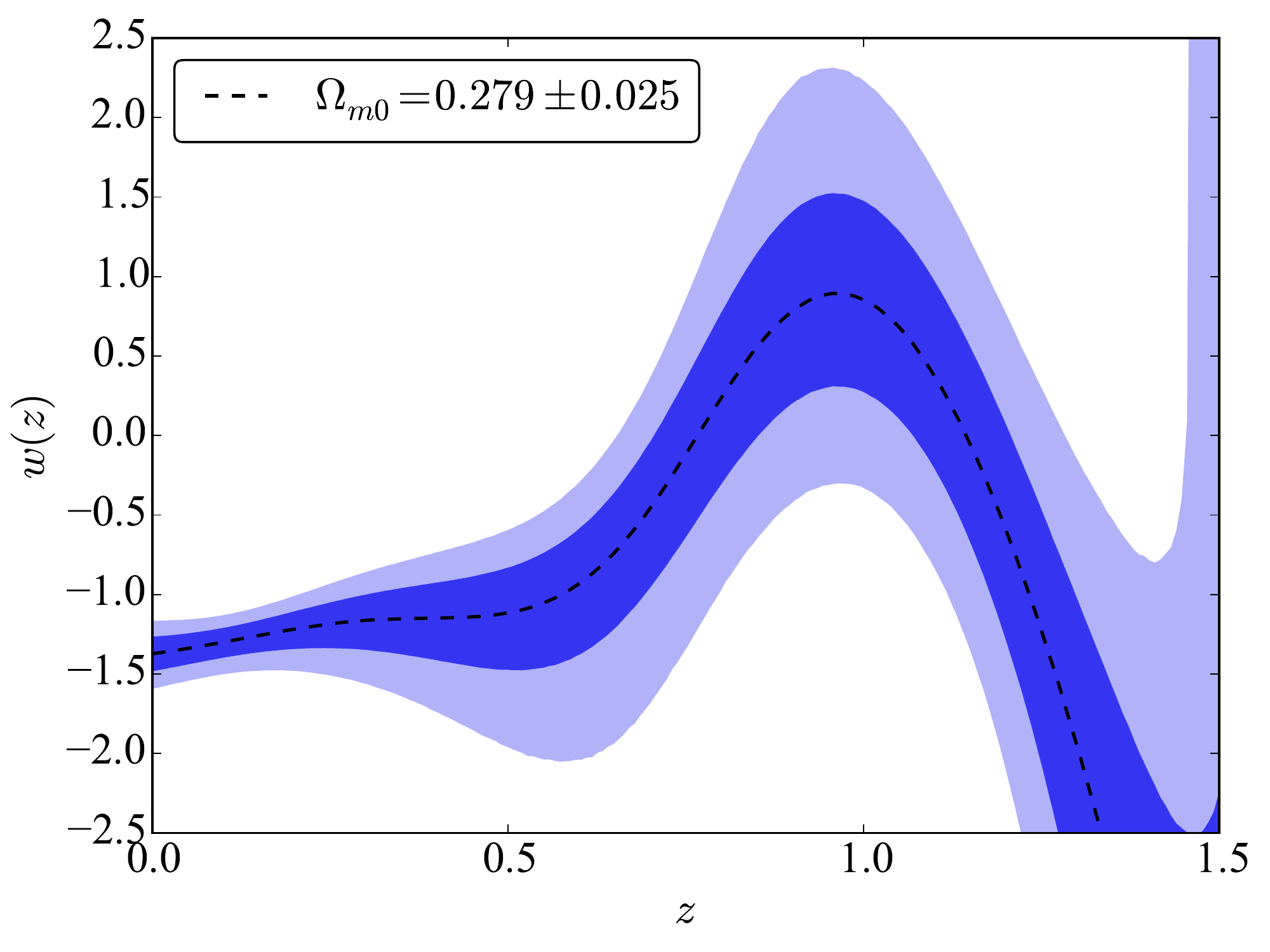}
    \includegraphics[width=5.9cm,height=5.4cm]{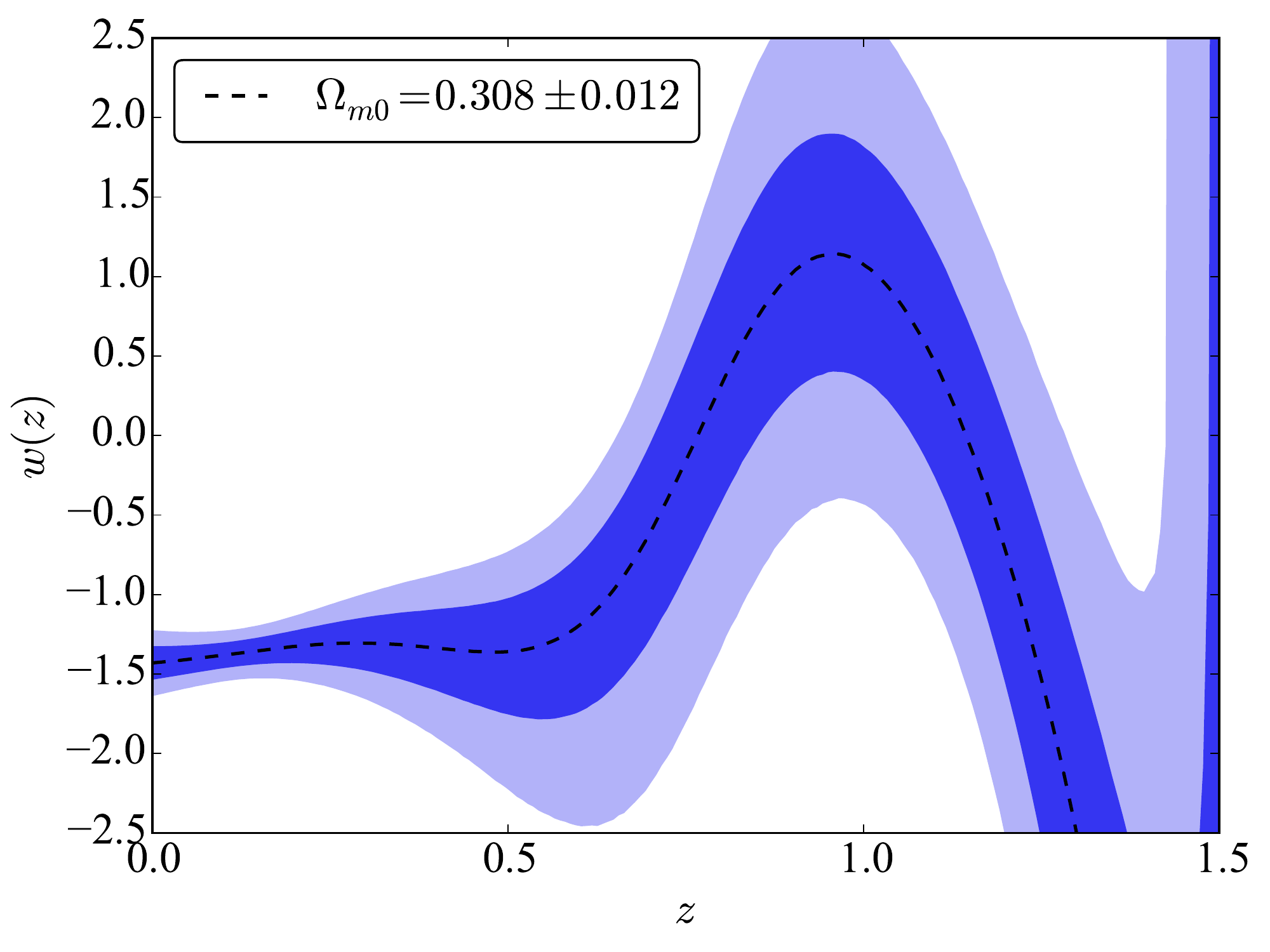}
    \caption{\label{JLA_w_om} Test of the effect of matter density parameter $\Omega_{m0}$ on $w$ reconstruction from the combination of JLA and $H(z)$ data. }
\end{figure*}

\begin{figure}
\centering
    \includegraphics[width=5.9cm,height=5.4cm]{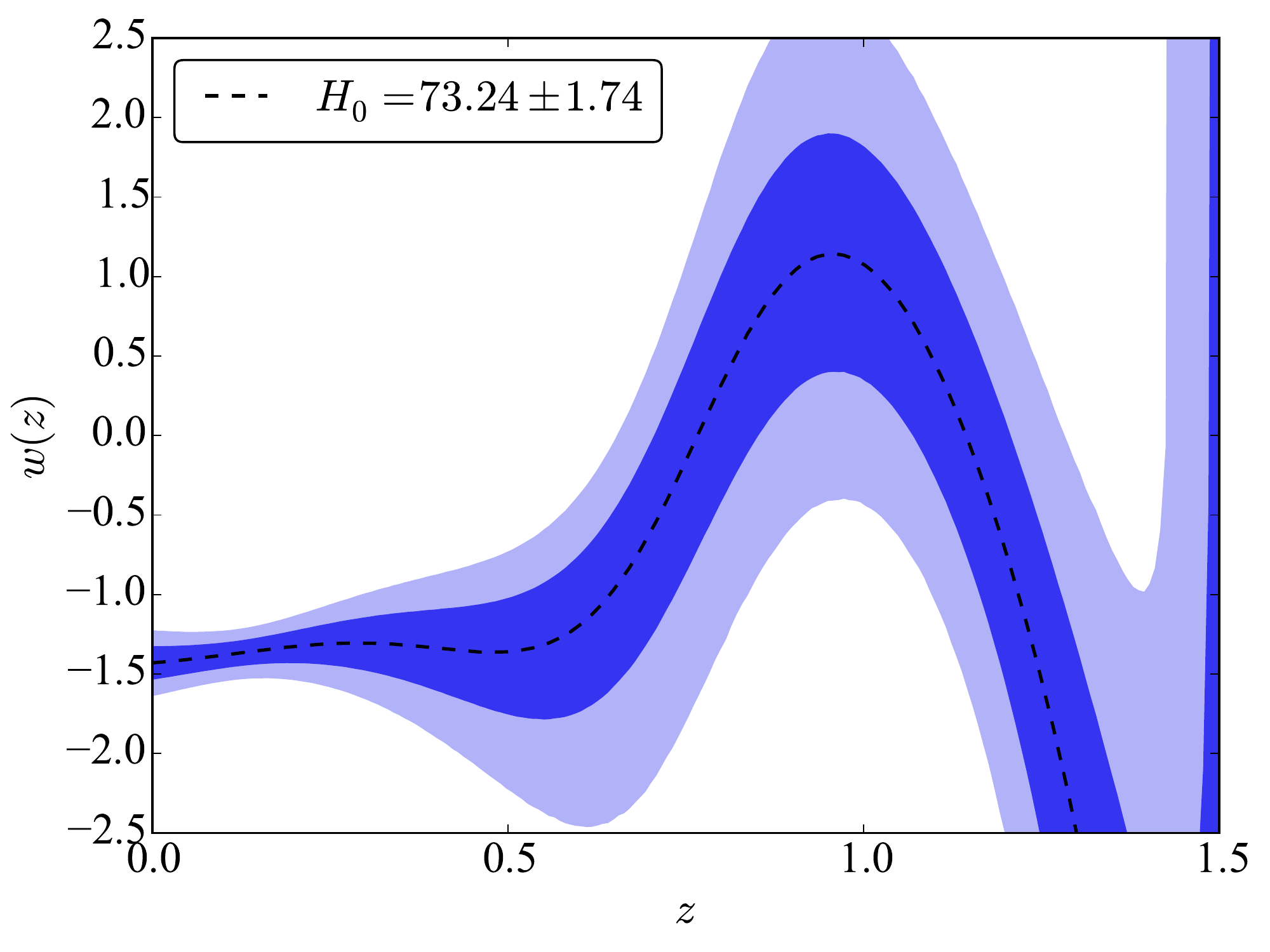}
     \includegraphics[width=5.9cm,height=5.4cm]{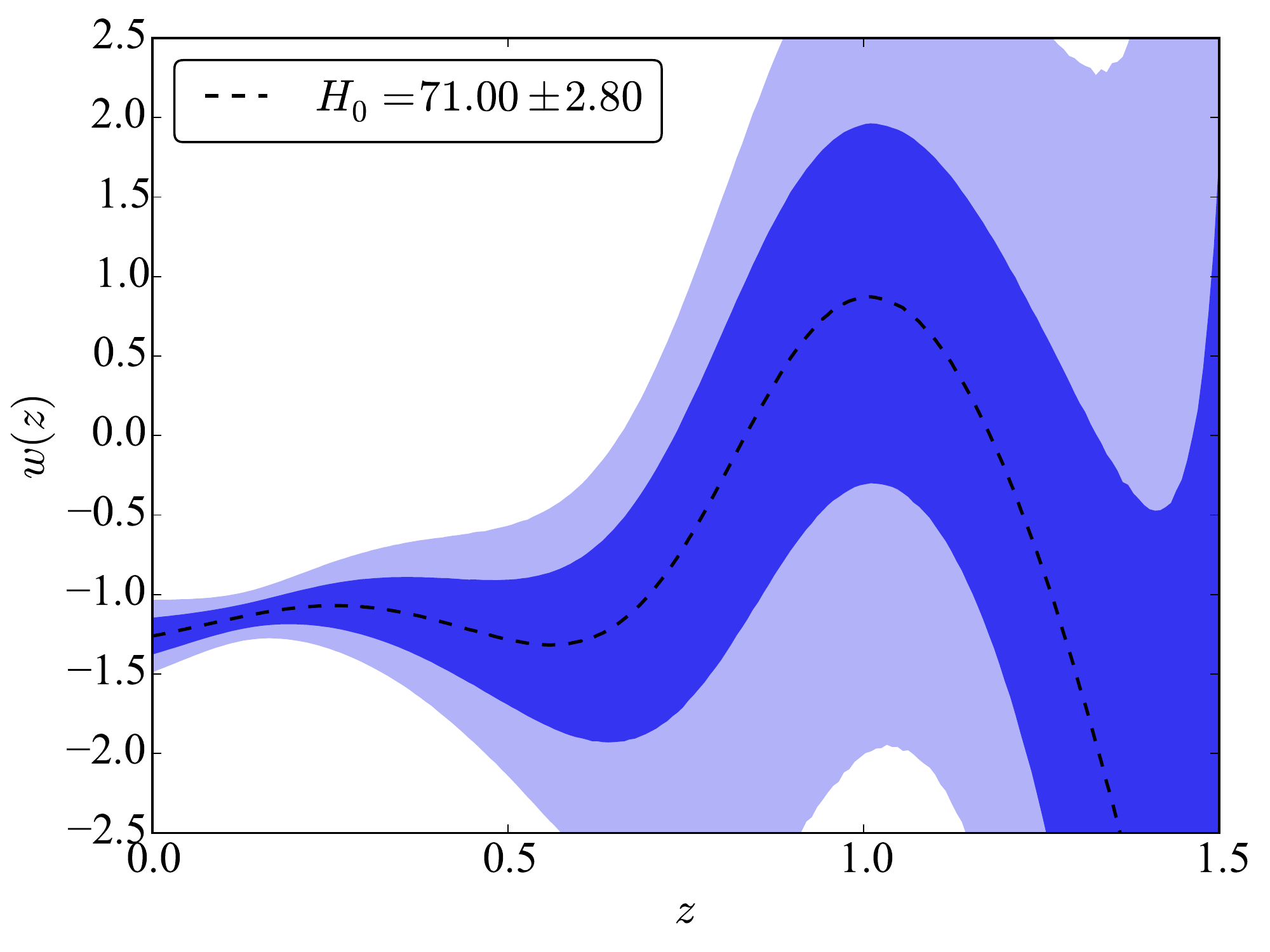}
    \caption{\label{JLA_w_H0} Test of the effect of Hubble constant $H_0$ on $w$ reconstruction from the combination of JLA and $H(z)$ data. }
\end{figure}

For the RSD data, they are in fact effects due to the differences between the observed distance and true distance on the galaxy distribution in redshift space. These differences are caused by the velocities in the overdensities deviation from the cosmic smooth Hubble flow expansion. Anisotropy of the radial direction  relative to transverse direction in the clustering of galaxies is correlated with the cosmic structure growth. Smaller deviation from the General Relativity implies a smaller anisotropic distortion in the redshift space. The RSD data is a very promising probe to distinguish the cosmological models, because different cosmological models may have similar background evolution, but cosmic growth of the structure in their models can be very distinct. Till now, the RSD data have been used extensively in previous literatures.  In this paper, we utilize the most recent RSD data from 2dF, 6dF, BOSS, GAMA, WiggleZ, galaxy surveys. We also consider the four very recent measurements from the eBOSS DR14 data \cite{2018arXiv180103043Z}. We collect the compilation in Table \ref{tab:fsig8_data}, which includes the survey, RSD data with errors, the corresponding references and year.

For the reconstruction using RSD, the Eq. \eqref{delta_end} indicates that an integral should be calculated to obtain the perturbation $\delta(z)$. Moreover, covariance of the $f \sigma_8$ should be propagated into the uncertainty of  $\delta(z)$. However, covariance propagation in the integral at all times is a difficult thing. For a simplicity, we only consider the uncertainties of $f \sigma_8$. For a Gaussian error propagation, the uncertainty of Hubble parameter can be expressed as
\be
\sigma_{E^2}^2 = \big( \frac{\partial E^2}{\partial \delta'} \big)^2 \sigma_{\delta'}^2 + \big( \frac{\partial E^2}{\partial I} \big)^2 \sigma_{I}^2 ,
\ee
where function $I = \int_0^{z} \frac{\delta}{1+z} (-\delta ') d z $. Finally, uncertainty of the EoS can be calculated via
\be
\sigma_w^2 = \big( \frac{\partial w}{\partial E^2} \big)^2 \sigma_{E^2}^2 + \big( \frac{\partial w}{\partial E^2 (z)' } \big)^2 \sigma_{E^2 (z)'}^2 .
\ee
In fact, it only influences the uncertainty of reconstruction. It does not impact the mean values of reconstruction of $w$. From the following $w$ reconstruction, we find that it is reasonable to do this manipulation.

\section{Result}
\label{result}

\begin{figure}
\includegraphics[width=9.2cm,height=5cm]{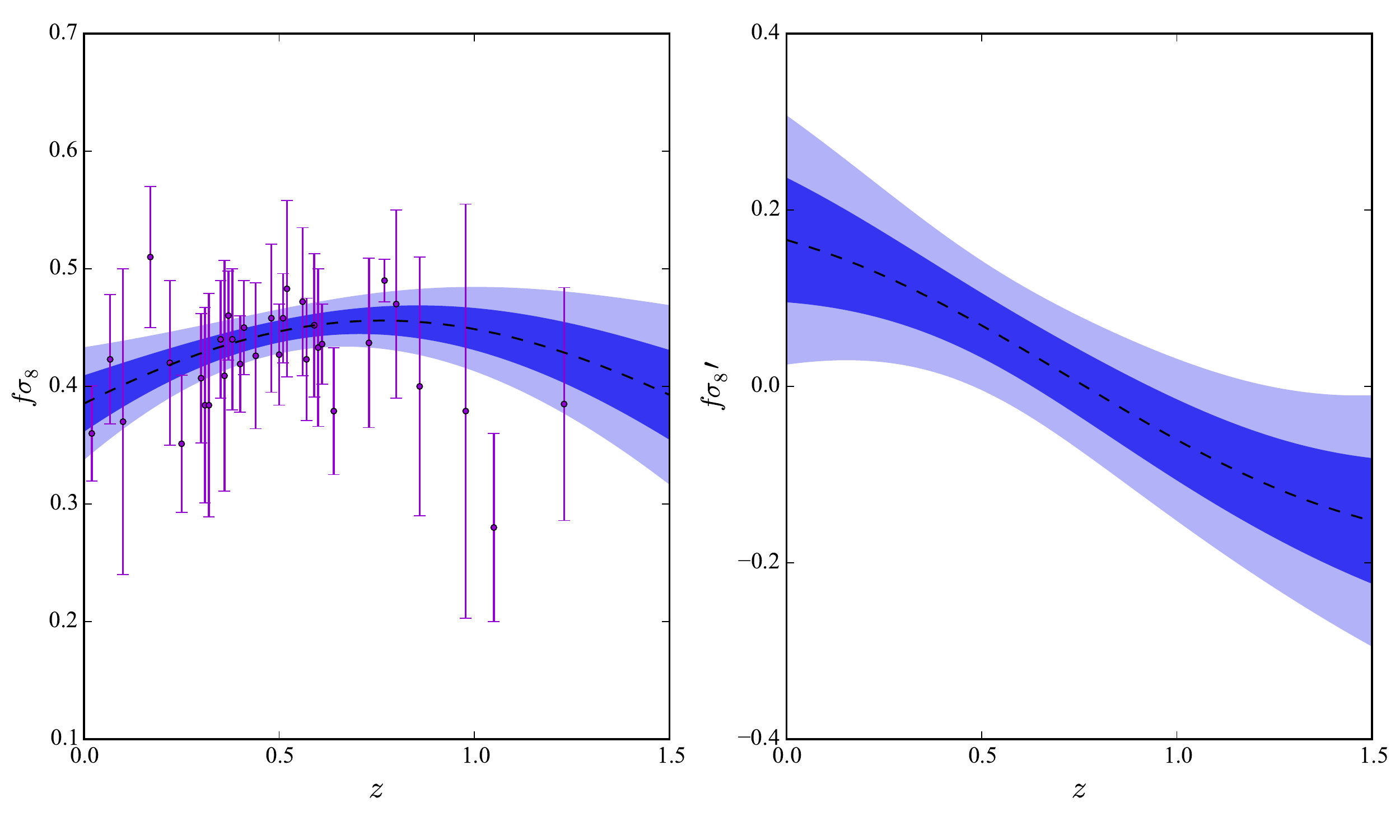}
    \caption{\label{RSD_D} The reconstruction of growth rate $f\sigma_8$ and its derivative using the RSD data. }
\end{figure}

To map the $w(z)$ of dark energy, we can reconstruct it from the Eq. \eqref{EoS:SN} for background data,  and Eq. \eqref{EoS:delta} for the perturbation data.

We report corresponding GP reconstruction in this section. In order to test the influence of some parameters on the reconstruction, we consider the effect of dark matter density parameter $\Omega_{m0}$ and Hubble constant $H_0$ for the background data, and effect of initial value $\delta (z=0)$, $\sigma_8 (z=0)$ and parameter  $\Omega_{m0}$ for the RSD data. For the matter density parameter $\Omega_{m0}$, we consider it in three cases: $\Omega_{m0}=0.25$, $0.279 \pm 0.025$ from WMAP-9 \cite{hinshaw2012nine} and $0.308 \pm 0.012$ from Planck 2015 \cite{ade2016planck}. For the initial value $\delta (z=0)$, we consider it as $\delta (z=0)=1$ and $0.7837$ in fiducial cosmology from Planck 2015 \cite{ade2016planck}. For the parameter $\sigma_8 (z=0)$, we consider it as $\sigma_8 (z=0)=0.821 \pm 0.023$ from WMAP-9 \cite{hinshaw2012nine} and $0.8149 \pm 0.0093$ from Planck 2015 \cite{ade2016planck}.

\subsection{Test on the Hubble parameter}
\label{test_Hz}

Before the reconstruction of $w$, we can perform a test on the Hubble parameter, in order to test whether they present a consistent background information.

In Fig. \ref{Fig:E2}, we plot the Hubble parameter reconstruction from the background data and perturbation data. From the upper panel, we find that Hubble parameter from the Union2.1+$H(z)$ data at low redshift is consistent with that from the JLA+$H(z)$. For high redshift, they present a different Hubble parameter, which indicates that they may give a slightly different $w$ reconstruction.

For the perturbation data, we find that they give a quite different $E^2(z)$ for different parameter $\Omega_{m0}$. On the one hand, the perturbation data present a smaller Hubble parameter $E^2$ at high redshift, when compared with the background data. Therefore, $w$ reconstruction from the perturbation data may be different from the background data. On the other hand, we note that the square of Hubble parameter $E^2(z)$ for different parameter $\Omega_{m0}$ is also different from each other. Especially, Hubble parameter $E^2(z)$ for $\Omega_{m0}=0.308 \pm 0.012$ at high redshift is negative, which indicates that the parameter $\Omega_{m0}$ should not be too bigger. Thus, we deem that the perturbation data are sensitive to the parameter $\Omega_{m0}$. They should be able to provide a tighter constraint on parameter $\Omega_{m0}$. Of course, it may be also highly influenced by the parameter $\Omega_{m0}$.

In short, we find a tension between the Hubble parameter from background data and perturbation data. And this tension may influence the dark energy reconstruction, leading to a different $w$.

\subsection{Reconstruction from the Union2.1 and $H(z)$ data}
\label{resultUnion21}

\begin{table}
\caption{\label{tab: EoS} Current EoS of dark energy $w_0$ at different cases for different observational data. }
\begin{tabular}{ccc}
\hline
\hline
$\Omega_{m0}$   & \qquad \qquad \qquad \qquad &  Union2.1+$H(z)$        \\
\hline
$\Omega_{m0} = 0.25$      & &  $w_0 = -1.1433 \pm 0.1460$       \\
WMAP-9 prior     & &  $w_0 = -1.1898 \pm 0.1560$      \\
Planck 2015 prior      &  &  $w_0 = -1.2393 \pm 0.1588$       \\
\hline
\nonumber \\
$\Omega_{m0}$   & \qquad \qquad \qquad \qquad &  JLA+$H(z)$        \\
\hline
$\Omega_{m0} = 0.25$      & &  $w_0 = -1.3196 \pm 0.0942$       \\
WMAP-9 prior     & &  $w_0 = -1.3725 \pm 0.1075$      \\
Planck 2015 prior      &  &  $w_0 = -1.4301 \pm 0.1050$       \\
\hline
\nonumber \\
$H_0$   & \qquad \qquad \qquad \qquad &  JLA+$H(z)$        \\
\hline
$H_0=73.24 \pm 1.74$      & &  $w_0 = -1.4297 \pm 0.1046$       \\
$H_0=71.00 \pm 2.80$    & &  $w_0 = -1.2612 \pm 0.1158$      \\
\hline
\nonumber \\
$\delta (z=0)$       &  &  RSD    \\
\hline
normalization value       & &  $w_0 = -0.8745 \pm 0.2490$       \\
fiducial value      & &  $w_0 = -0.8745 \pm 0.2490$       \\
\hline
\nonumber \\
$\sigma_8 (z=0)$        & &  RSD    \\
\hline
WMAP-9 prior      & &  $w_0 = -0.8653 \pm 0.2496$       \\
Planck 2015 prior      & &  $w_0 = -0.8980 \pm 0.2477$        \\
\hline
\nonumber \\
$\Omega_{m0}$    &  &  RSD        \\
\hline
$\Omega_{m0} = 0.25$      & &  $w_0 = -0.6433 \pm 0.2430$       \\
WMAP-9 prior     & &  $w_0 = -0.9490 \pm 0.2522$      \\
Planck 2015 prior      & &  $w_0 = -0.9026 \pm 0.2502$       \\
\hline
\hline
\end{tabular}
\end{table}

We show the GP reconstructions for combination data in Figs. \ref{union21_D} and \ref{union21_w}. The dashed lines and shaded region correspond to the mean values and errors of reconstructions, respectively.

We find that the reconstructions are consistent with the observational data, as shown in the first two panels of Fig. \ref{union21_D}. We also note that $D'$ and $D''$ in this figure are quite different from the previous work using the supernova data alone. For the supernova data alone, their derivatives  $D'$ and $D''$ change smoothly and softly. In contrast, the derivatives change acutely in this paper. This is because the input $H(z)$ data as a prior of the derivative $D'$ change the covariance function $k(z, \tilde{z})$ at different points. Thus, they present a different GP reconstruction.

To test the impact of matter density parameter $\Omega_{m0}$, we plot the $w(z)$ reconstruction for different $\Omega_{m0}$ in Fig. \ref{union21_w}. Comparison shows that the parameter $\Omega_{m0}$ produces slight influence on the reconstruction. First, they present a similar estimation of dark energy  $w(z)$ over the redshift. Current EoS of dark energy at all cases is $w_0 <-1$ at 68\% C.L., as shown in Table \ref{tab: EoS}. Second, the cosmological constant model with $w=-1$ at 95\% C.L. cannot be ruled out by the background observational data, especially for high redshift. Third, mean values of reconstruction hint a dynamical dark energy, but they cannot rule out the non-evolution model at 95\% C.L. Last, uncertainty of reconstruction at redshift $z>0.5$ has becomes very large. It indicates that a precise evaluation on $w(z)$ at high redshift is still a luxury from current background data.

\begin{figure*}
\centering
\includegraphics[width=18cm,height=6cm]{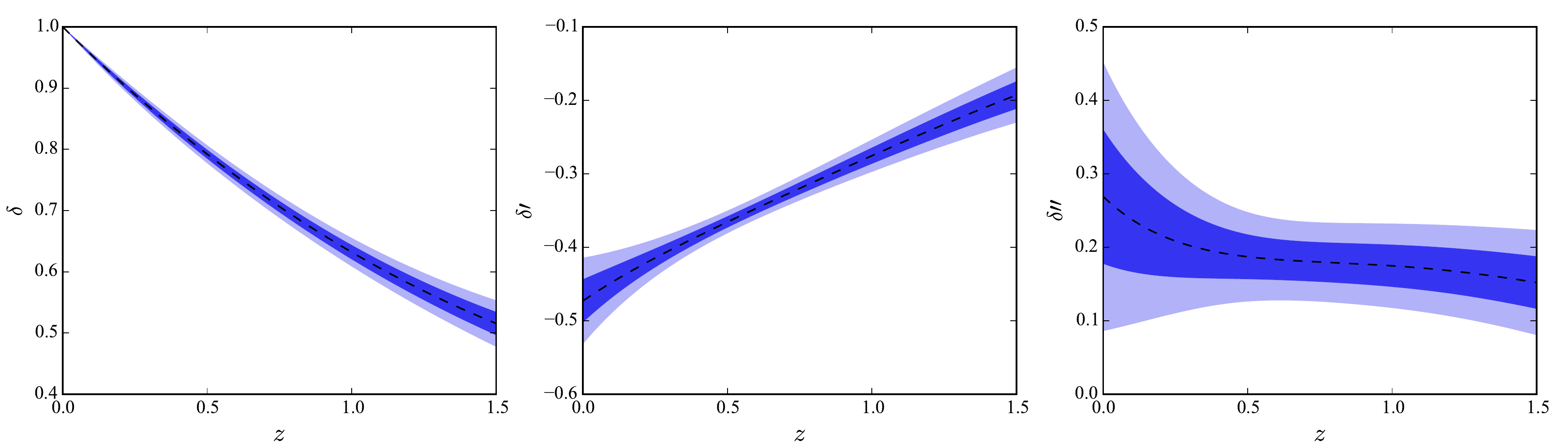}
    \caption{\label{RSD_delta1} The reconstruction of perturbation $\delta$ and its derivatives with initial condition $\delta(z=0)=1$ and $\sigma_8 (z=0)=0.8149$ using the RSD data. }
\end{figure*}

\subsection{Reconstruction from the JLA and $H(z)$ data}
\label{resultJLA}

We test the influence of parameter $\Omega_{m0}$ and $H_0$ on the $w$ reconstruction from JLA and $H(z)$ data.

\subsubsection{Effect of the parameter $\Omega_{m0}$}
\label{JLA_om}

We show the test of parameter $\Omega_{m0}$ on $w$ reconstruction from JLA and $H(z)$ data in Fig. \ref{JLA_w_om}. From the comparison with Fig. \ref{union21_w}, we find that this combination presents a similar $w$ reconstruction as the Union2.1 and $H(z)$ data. For low redshift, EoS $w < -1$ can be highlighted, which can be seen from the $w_0$ in Table \ref{tab: EoS}. We also note that errors of $w$  from the JLA combination is smaller. This is because the JLA data have more samples with high precision, which can present a tighter constraint. Same as the Union2.1 data combination, the JLA combination also hint a dynamical dark energy. However, because of its smaller errors, the cosmological constant model is not so consistent with the reconstruction from JLA and $H(z)$ data.

\subsubsection{Effect of the parameter $H_0$}
\label{JLA_H0}

We present the test of influence of different Hubble constant in Fig. \ref{JLA_w_H0}. Firstly, we find that JLA data combination in this case give a similar $w$ as above reconstruction. A dynamical $w$ is also presented. Secondly, The Hubble constant has a notable influence on the $w$, as shown in Table \ref{tab: EoS} on the current EoS $w_0$. Moreover, errors of $w$ in the upper panel is smaller, due to its smaller uncertainty of Hubble constant. The cosmological constant model in the lower panel cannot be distinguished from the reconstruction.

\subsection{Reconstruction from the RSD data}
\label{resultRSD}

Using the GP method, we obtain the current growth rate $f\sigma_8 (z=0) = 0.3854 \pm 0.0239$ and $f\sigma_8' (z=0) = 0.1660 \pm 0.0706$, as shown in Fig. \ref{RSD_D}. We should emphasize that this estimation about the current growth rate is model-independent. We also note that the growth rate was decreasing. Within 95\% C.L., the derivative $f\sigma_8' $ gradually decreases to nagative value for redshift $z \gtrsim 0.5$.

For the perturbation $\delta$ and EoS $w(z)$, they are dependent of the initial values $\delta (z=0)$, $\sigma_{8} (z=0)$ and matter density parameter $\Omega_{m0}$. In the following text, we intend to test their influences on the reconstruction.

\subsubsection{Effect of the parameter $\delta (z=0)$}
\label{effect_delta0}
\begin{figure}
\centering
\includegraphics[width=7cm,height=5.5cm]{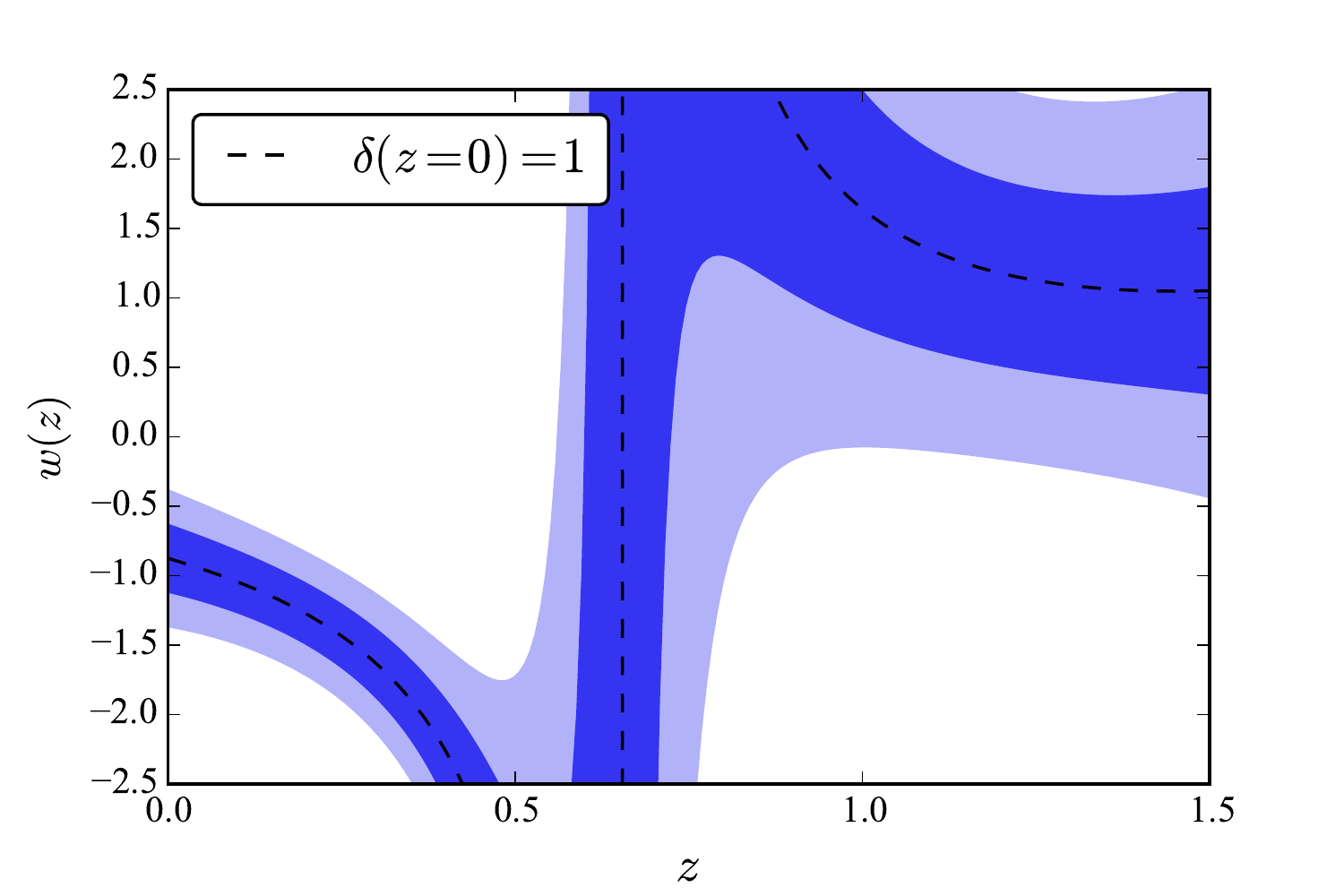}
\includegraphics[width=7cm,height=5.5cm]{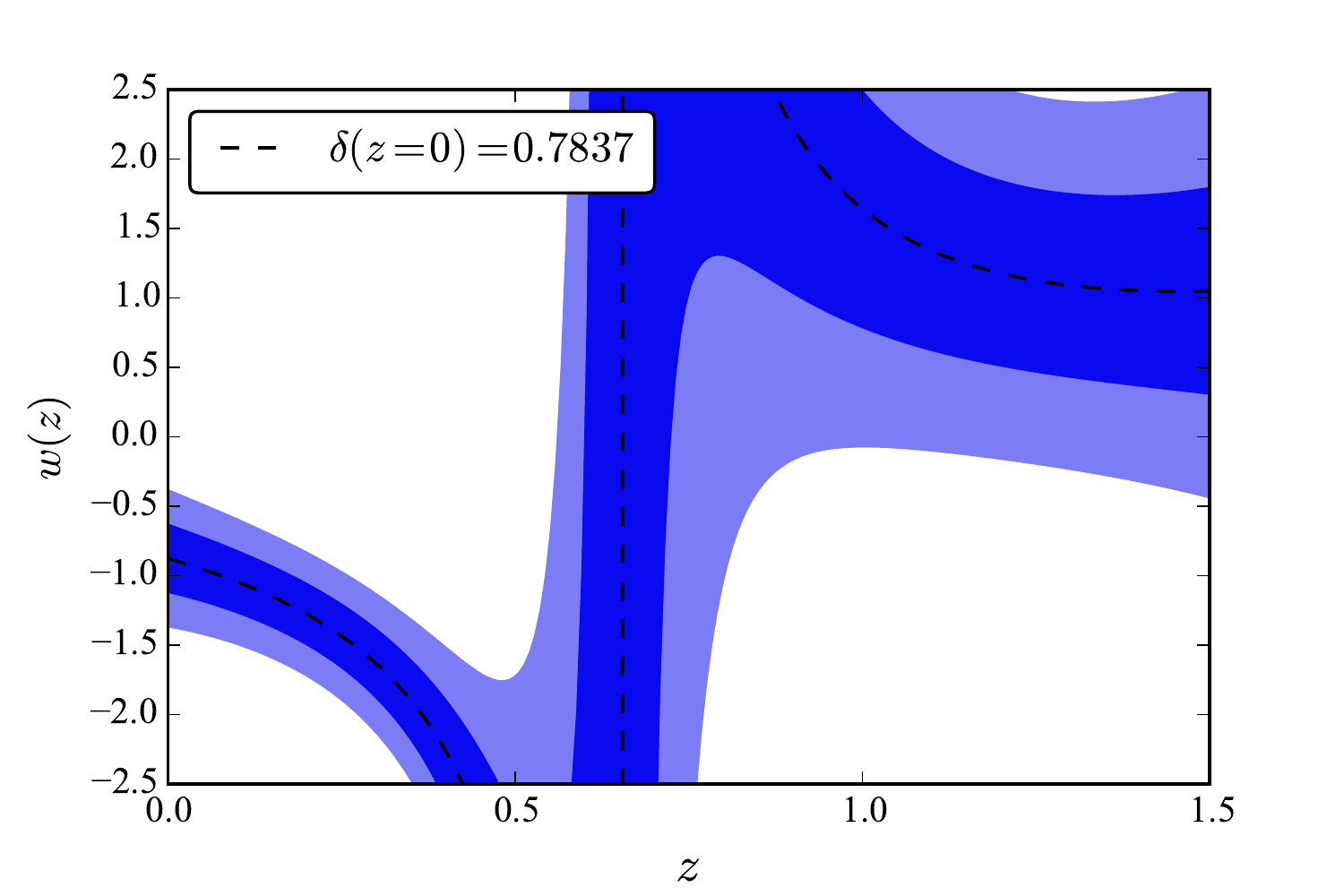}
    \caption{\label{Fig:delta_w} Test of the effect of initial condition $\delta(z=0)$ on $w$ reconstruction from the RSD data.  }
\end{figure}

The initial value $\delta (z=0)$ is unknown for us. Generally, it is taken as the normalization value $\delta (z=0)=1$ \cite{2017JCAP...08..008G,2016arXiv161202484C} or a fiducial value. Using the GP approach, we plot the related reconstructions in Fig. \ref{RSD_delta1}. Assuming the constant $\sigma_8 (z=0)=0.8149$, we reconstruct the perturbation $\delta$ and its derivatives. We find that $\delta$ decreases with the increasing redshift. To higher redshift, it may decrease to zero. For the $\delta '$, investigation about it was absent. From the GP method, we obtain that $\delta '(z=0)=-0.4729 \pm 0.0294$ and  $\delta ''(z=0)=0.2691 \pm 0.0915$. The first derivative $\delta '$ with 95\% C.L. is negative, although it increases with the redshift.

To test the influence of different initial values $\delta (z=0)$, we reconstruct the dark energy $w$ at two cases, normalization value and fiducial value $\delta (z=0)=0.7837$, under the same assumption of matter density parameter $\Omega_{m0}=0.308$. We plot them in Fig. \ref{Fig:delta_w}. From the comparison, we find that they both present a dynamical $w$, almost without any difference. This reconstruction is much different from the background data. From the list in Table \ref{tab: EoS}, we find that the current EoS of dark energy $w_0$ in two cases are the same. It indicates that the initial values $\delta (z=0)$ has no influence on the reconstruction. Therefore, it is safe for us to take the initial value $\delta (z=0)=1$ in future calculation.

\subsubsection{Effect of the parameter $\sigma_{8} (z=0)$}
\label{effect_sig80}
\begin{figure}
\centering
\includegraphics[width=7cm,height=5.5cm]{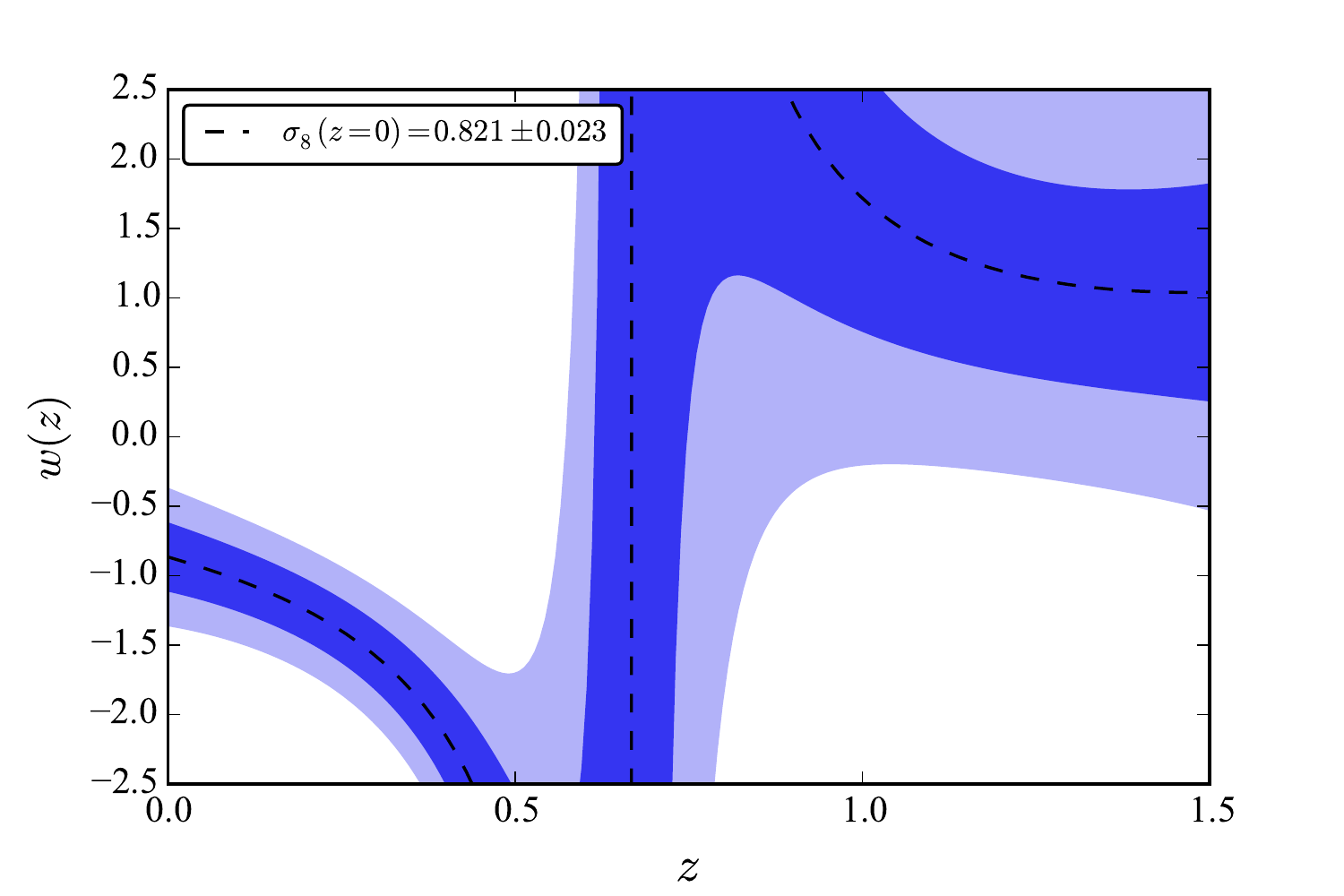}
\includegraphics[width=7cm,height=5.5cm]{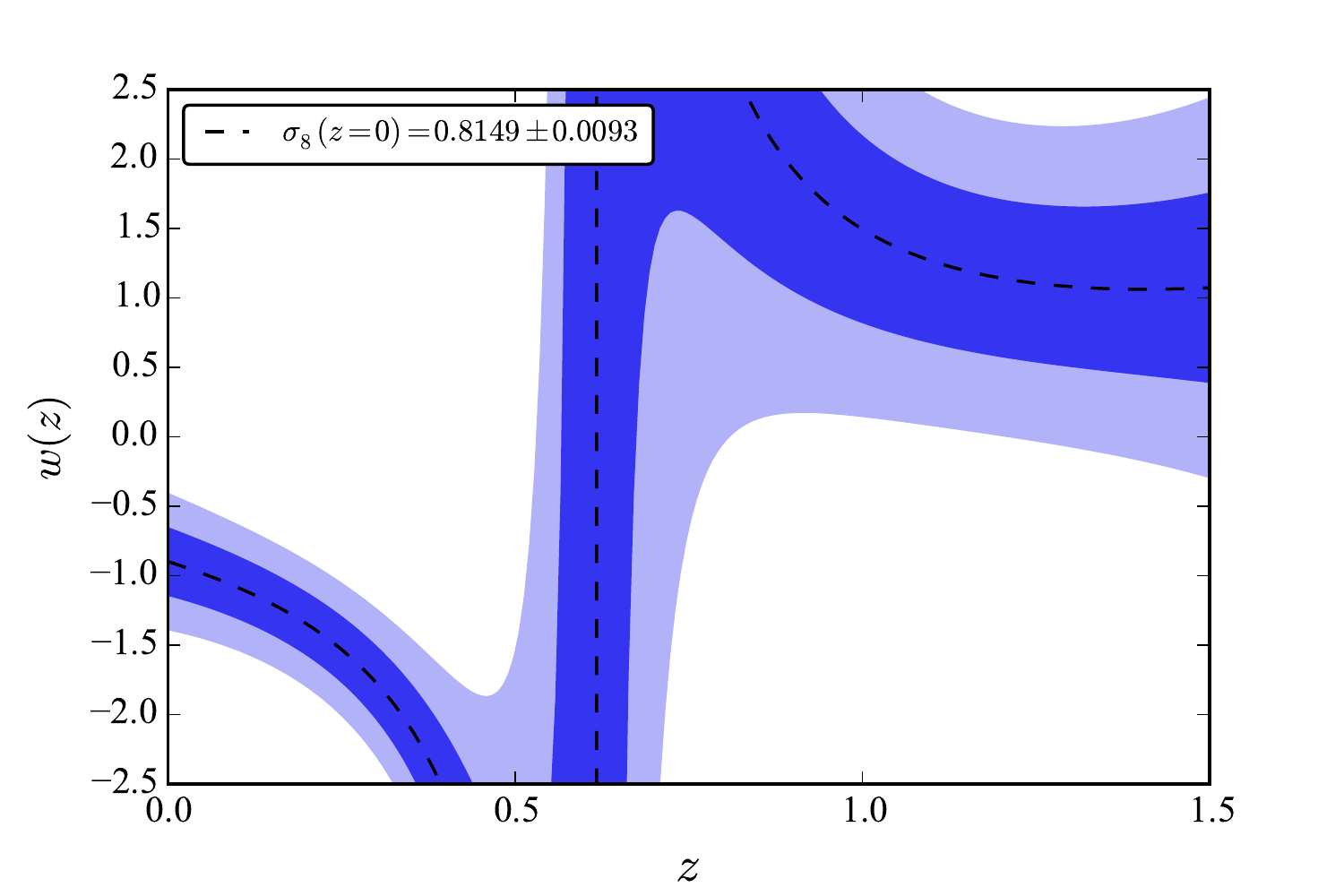}
    \caption{\label{Fig:sigma8_w} Test of the effect of initial condition $\sigma_8(z=0)$ on $w$ reconstruction from the RSD data.  }
\end{figure}

To test the influence of $\sigma_{8} (z=0)$, we respectively consider its value in the WMAP-9 prior and Planck 2015 prior. The initial value and density parameter are fixed to be $\delta (z=0)=1$ and $\Omega_{m0}=0.308$. We show the reconstruction in Fig. \ref{Fig:sigma8_w}. From the comparison, firstly, we find that the RSD data in these cases also present a dynamical $w$, similar as the constraint in Fig. \ref{Fig:delta_w}. Moreover, the cosmological constant model can be demonstratively distinguished from the reconstruction. Secondly, we find that current EoS of dark energy are $w_0 = -0.8653 \pm 0.2496$ for the WMAP-9 prior and  $w_0 = -0.8980 \pm 0.2477$ for the Planck 2015 prior, respectively. It indicates that the initial value $\sigma_{8} (z=0)$ can influence the $w$  reconstruction slightly, which can be evidenced in Table \ref{tab: EoS}.

\begin{figure*}
    \begin{center}
\includegraphics[width=5.9cm,height=5.5cm]{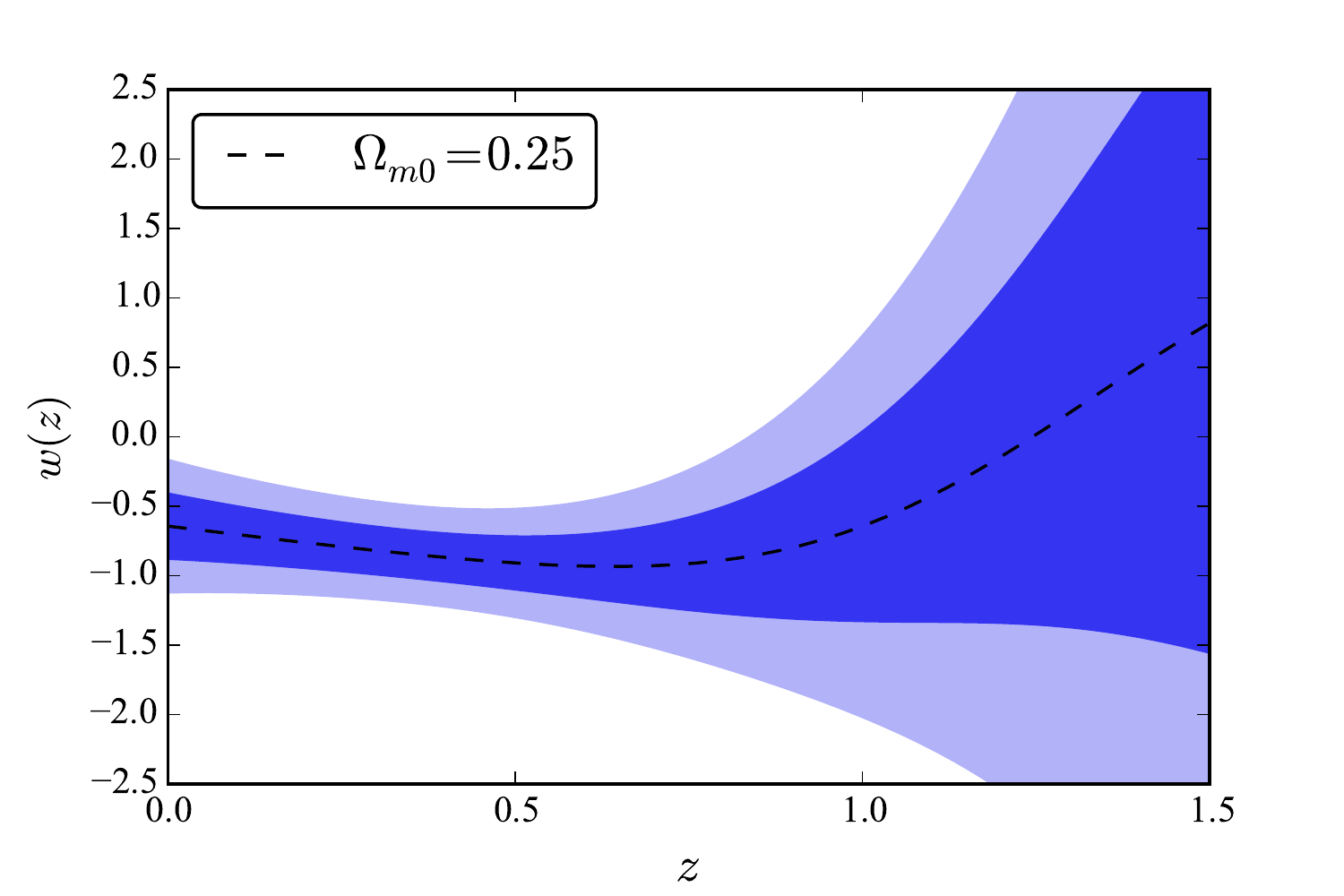}
\includegraphics[width=5.9cm,height=5.5cm]{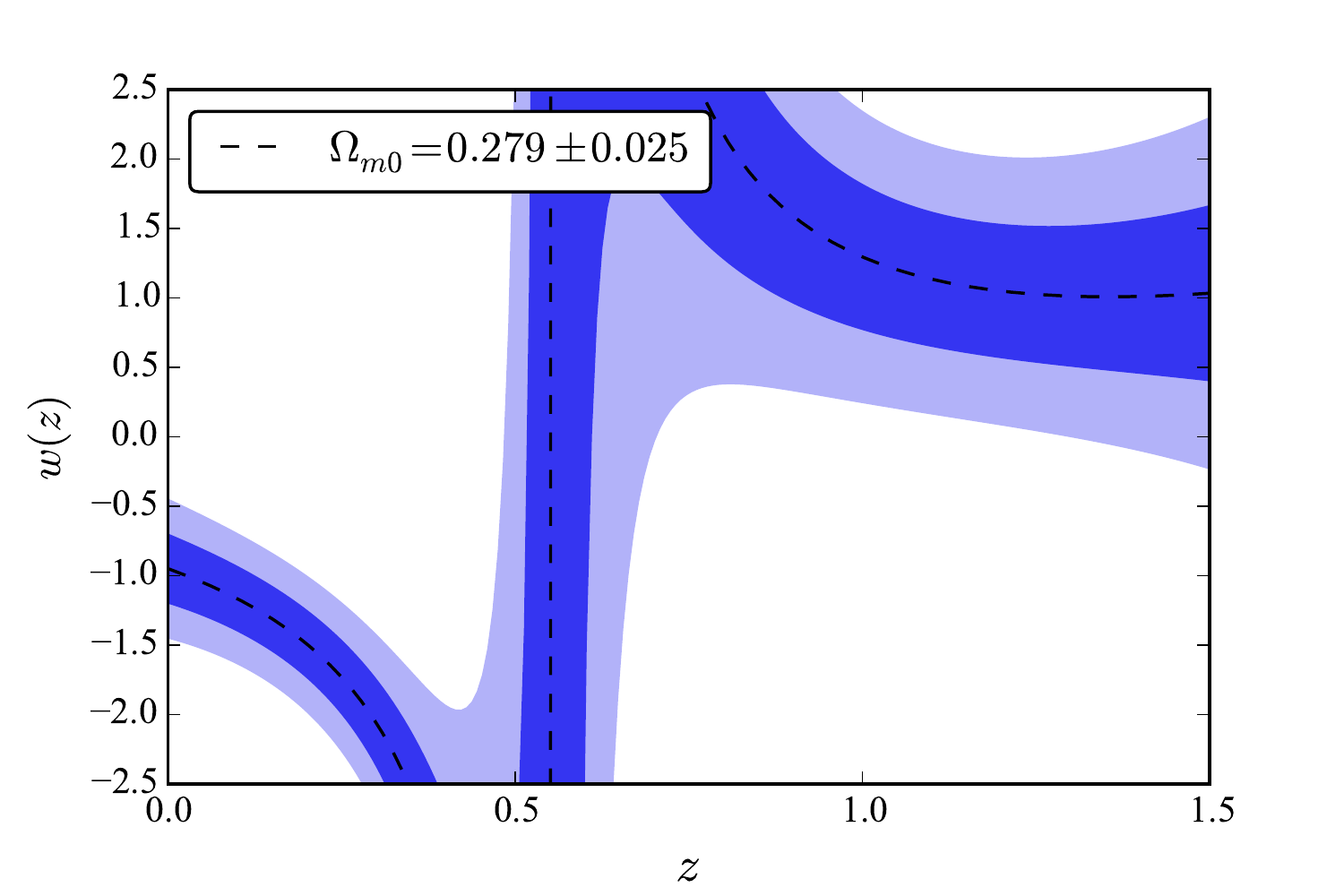}
\includegraphics[width=5.9cm,height=5.5cm]{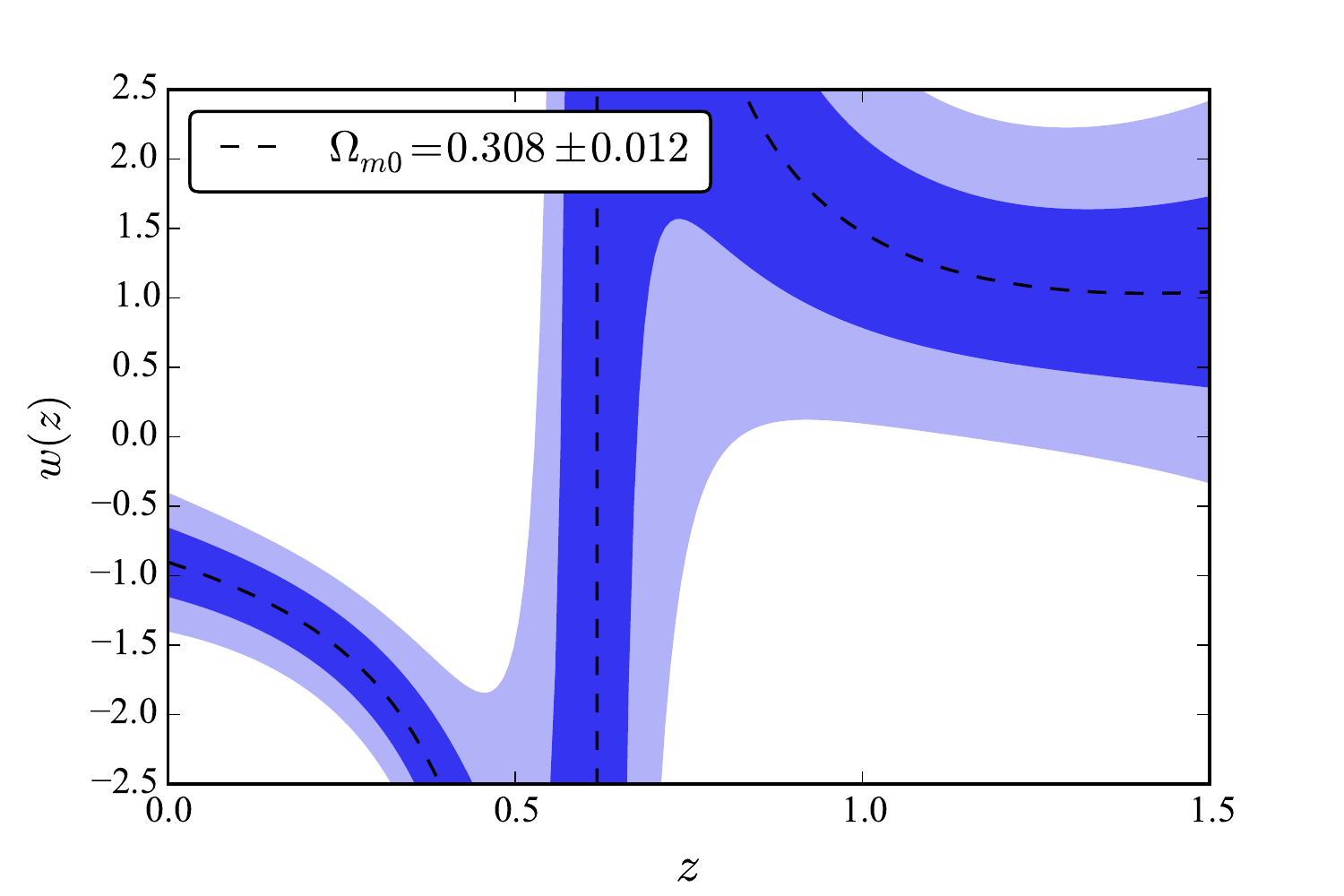}
    \end{center}
    \caption{\label{Fig:om_w} Test of the effect of matter density parameter $\Omega_{m0}$ on $w$ reconstruction from the RSD data. }
\end{figure*}

\subsubsection{Effect of the parameter $\Omega_{m0}$}
\label{effect_om}

For the background data, we consider three types of matter density parameter $\Omega_{m0}$ on the reconstruction. It implies that the parameter $\Omega_{m0}$ does not produce drastic effect on the $w$.

In this subsection, we also test this effect for different parameter $\Omega_{m0}$. By fixing the initial value $\delta (z=0)=1$ and  $\sigma_{8} (z=0)=0.8149$, we perform this test, and make a series of comparisons in Fig. \ref{Fig:om_w}. Firstly, we find that effect of matter density parameter $\Omega_{m0}$ is very significant. Specifically, in the first panel, RSD data for $\Omega_{m0}=0.25$ present a very different $w$, when compared with two other reconstructions. In this case, the reconstruction favors dark energy with a constant $w$. Moreover, the cosmological constant model cannot be ruled out. However, we also note that this $w$ is different from the one obtained by the background data, even though the latter also cannot exclude the cosmological constant model. We think that the reason can be understood from the Hubble parameter in Fig. \ref{Fig:E2}. From the comparison in that figure, we find that the RSD data present a quite different Hubble parameter at three different parameter $\Omega_{m0}$, which implies that RSD data are sensitive to the parameter $\Omega_{m0}$. From the definition of $w$ in Eq. \eqref{EoS:delta}, we note that it inevitably depends on the Hubble parameter. Therefore, it should not difficult to understand why the $w$ reconstruction for different parameter $\Omega_{m0}$ are different. We also examine the definition of $w$ in Eqs. \eqref{EoS:SN} and \eqref{EoS:delta}. We find that denominator of Eq. \eqref{EoS:delta} crosses the zero, when the parameter $\Omega_{m0} \gtrsim 0.30$. While for the background data, it happens only when parameter  $\Omega_{m0} \gtrsim 0.45$. Therefore, a big steep slop in $w$ from perturbation data is presented. Secondly, the $w$ in two other cases is also suggested to be a dynamical one. This is same as the reconstructions in above several scenarios. The cosmological constant model is highly discordant with these model-independent reconstructions.

In short, the RSD data are strongly influenced by the matter density parameter $\Omega_{m0}$, and can present a dynamical $w$ reconstruction, which is different from the reconstruction using background data.

\section{Conclusion and discussion}
\label{conclusion}

In this paper, we carry out a model-independent reconstruction on the dark energy $w$, using the Gaussian processes approach. The observational data we use are supernova data, $H(z)$ parameter and growth rate data.

Different from previous work, the background data we use are the combination of Union2.1 supernova data,  JLA data and $H(z)$  data. The $H(z)$ parameter data here are used together with the supernova data, and as the derivative of distance $D$. Their inclusion can provide more information about the cosmic evolution. Moreover, we find that inclusion of them can significantly change the cosmic evolution, as the reconstruction in Fig. \ref{union21_D}.

Using the combination of supernova and $H(z)$ data, we find that the background data present a hint of dynamical $w$. However, it cannot exclude the $\Lambda$CDM model within 95\% C.L.. Recently, Zhao et al. \cite{zhao2017dynamical} investigated the dark energy using the latest data including CMB temperature and polarization anisotropy spectra, supernova, BAO from the clustering of galaxies and from the Lyman-$\alpha$ forest, Hubble constant and $H(z)$.  They found that the dynamical dark energy can relieve the Hubble constant tension and is preferred at a 3.5$\sigma$ significance level. Moreover, the upcoming dark energy survey DESI++ would be able to provide a decisive Bayesian evidence. Comparing with their reconstruction, this work presents a similar evolution on the $w(z)$, which is consistent with their determination. At the same time, we test the effect of matter density parameter $\Omega_{m0}$ and Hubble constant. The comparisons indicate that parameter $\Omega_{m0}$ has a slight influence on the $w$ reconstruction from background data. However, the Hubble constant presents a notable influence on the reconstruction.

Another work we have done is to investigate the dark energy using growth rate data at a perturbation level. We obtain $f\sigma_8 (z=0) = 0.3854 \pm 0.0239$ and $f\sigma_8' (z=0) = 0.1660 \pm 0.0706$, which are model-independent, as shown in Fig. \ref{RSD_D}. For the perturbation $\delta$, its objective estimation was absent. From the GP reconstruction, we obtain its derivatives $\delta '(z=0)=-0.4729 \pm 0.0294$ and  $\delta ''(z=0)=0.2691 \pm 0.0915$, assuming the initial value $\sigma_8 (z=0)=0.8149$.

We also test the effect of three parameters on the $w$ reconstruction from perturbation data. We find that the initial value $\delta (z=0)$ has no effect on the $w$ reconstruction. It is safe for us to take the normalization value $\delta (z=0)=1$. While for the parameter $\sigma_8 (z=0)$, it presents a slight influence on the reconstruction. However, importantly, the matter density parameter $\Omega_{m0}$ has a notable influence on the $w$ reconstruction. Similar as the background data, the growth rate data also provide a dynamical dark energy $w$. However, an obvious difference between them is that the perturbation data have a more promising potential to distinguish the $\Lambda$CDM model.

One improvement of our work concerns the GP method, a model-independent approach, allowing us to break the limitation of specific model. Another potential difference of our work is the extension of data types.
In previous analysis, luminosity distance and $H(z)$ data were often used alone. In this work, we not only use them together, but also compare their reconstruction with the perturbation data.

In addition, our consideration on the effect of some parameters also presents a full complement to previous study on this subject. Especially, we find that the matter density parameter $\Omega_{m0}$ has a notable influence on the perturbation data, but no important influence on the background data.

\section*{Acknowledgments}

We thank the anonymous referee whose suggestions greatly helped us improve this paper. H. Li is supported by the Youth Innovation Promotion Association Project of CAS. M.-J. Zhang is funded by China Postdoctoral Science Foundation under grant No. 2015M581173. The research is also supported in part by NSFC under Grant Nos. 11653001, Pilot B Project of CAS (No. XDB23020000) and Sino US Cooperation Project of Ministry of Science and Technology (No. 2016YFE0104700).

\bibliography{wGP}
\end{document}